\documentclass[twocolumn,superscriptaddress]{revtex4-2}
\usepackage{graphicx}
\usepackage{lipsum}
\usepackage{float}
\usepackage[utf8]{inputenc}
\usepackage{color}
\usepackage{xcolor}
\usepackage{dcolumn}
\usepackage{bm}
\usepackage{amsmath}

\renewcommand{\vec}[1]{\mathbf #1}

\begin{document}

\title{Active particles in tunable compressible environments}
\author{Venkata Manikantha Sai Ganesh Tanuku}
\email{vtanuku@uni-mainz.de}
\affiliation{Institute of Physics, Johannes Gutenberg University, 55128 Mainz, Germany}
\author{Isha Malhotra}
\affiliation{Institute for Theoretical Physics II, Heinrich-Heine-Universität Düsseldorf, Universitätsstr. 1, 40225 Düsseldorf, Germany}
\author{Lorenzo Caprini}
\affiliation{Institute for Theoretical Physics II, Heinrich-Heine-Universität Düsseldorf, Universitätsstr. 1, 40225 Düsseldorf, Germany}
\affiliation{Present address: Sapienza University of Rome, Piazzale Aldo Moro 2, Rome, Italy}
\author{Hartmut Löwen}
\affiliation{Institute for Theoretical Physics II, Heinrich-Heine-Universität Düsseldorf, Universitätsstr. 1, 40225 Düsseldorf, Germany}
\author{Thomas Palberg}
\email{palberg@uni-mainz.de}
\affiliation{Institute of Physics, Johannes Gutenberg University, 55128 Mainz, Germany}
\author{Ivo Buttinoni}
\email{Ivo.Buttinoni@hhu.de}
\affiliation{Institute for Experimental Physics of Condensed Matter, Heinrich-Heine-Universität Düsseldorf, Universitätsstr. 1, 40225 Düsseldorf, Germany}

\date{\today}

\begin{abstract}
Active particles affect their environment as much as the environment affects their active motion. Here, we present an experimental system where both can be simultaneously adjusted \textit{in situ} using an external AC electric field. The environment consists in a two-dimensional bath of colloidal silica particles, whereas the active particles are gold-coated Janus spheres. As the electric field orthogonal to the planar layer increases, the former become stiffer and the latter become faster. The active trajectories exhibit enhanced rotational motion where the reorientation frequency increases with the particle speed, an effect that culminates in a chiral active motion. We demonstrate that self-sustained reorientations arise from local compressions and interaction asymmetries, revealing a general particle-level mechanism where changes in the mechanical properties of the environment reshape active trajectories.  
\end{abstract}

\keywords{active Brownian motion, microswimmers, active matter}
\maketitle

\section{Introduction}\label{intro}

In recent years, synthetic active colloids have become an important model system to reproduce the behavior of biological microswimmers in complex environments. They are also promising for applications in microrheology, drug delivery, and micromachines~\cite{patra2013intelligent,howes2014colloidal,bechinger2016active,elgeti2015physics}. Their free active motion in simple fluids like water is well understood, with the dynamics primarily governed by viscous stresses and rotational diffusion~\cite{bechinger2016active}. However, complex environments significantly vary the ability of the particles to swim and reorient~\cite{bechinger2016active,ketzetzi2024active}. In particular, the persistence time of an active trajectory, {\sl i.e.} the average duration of its straight paths, is often affected by collisions with surfaces and obstacles~\cite{schaar2015detention,fins2024steer,volpe2017topography,brown2016swimming,spagnolie2015geometric,morin2017distortion}, alignments with external fields~\cite{lozano2016phototaxis,fernandez2020feedback,carrasco2023sedimentation}, flows~\cite{palacci2015artificial,sharan2022upstream} and interfaces~\cite{wang2015enhanced,wang2017Janus,dietrich2017two}. A similar scenario is encountered in biological microswimmers, such as bacteria and living cells, which often adapt their motion to external environmental stimuli~\cite{bechinger2016active}. 

The motion of microswimmers in complex fluids is another paradigmatic example of feedback between activity and environmental properties, with important consequences for self-transport, self-organization and active flows~\cite{patteson2016active,spagnolie2023swimming}. In particular, recent experiments and simulations reported a dramatic decrease of persistence time~\cite{gomez2016dynamics,qi2025unravel,saito2025self} and even a break of polar symmetry~\cite{narinder2018memory,decorato2025enhanced,decorato2021spontaneous} in polymer solutions and nanostructured media. Despite the abundance of complex fluids in nature, these feedback mechanisms in active matter remain largely unexplored and have not yet been harnessed to adjust \textit{in situ} the motion of active particles. 

\begin{figure}[t]
	\includegraphics[scale=0.7]{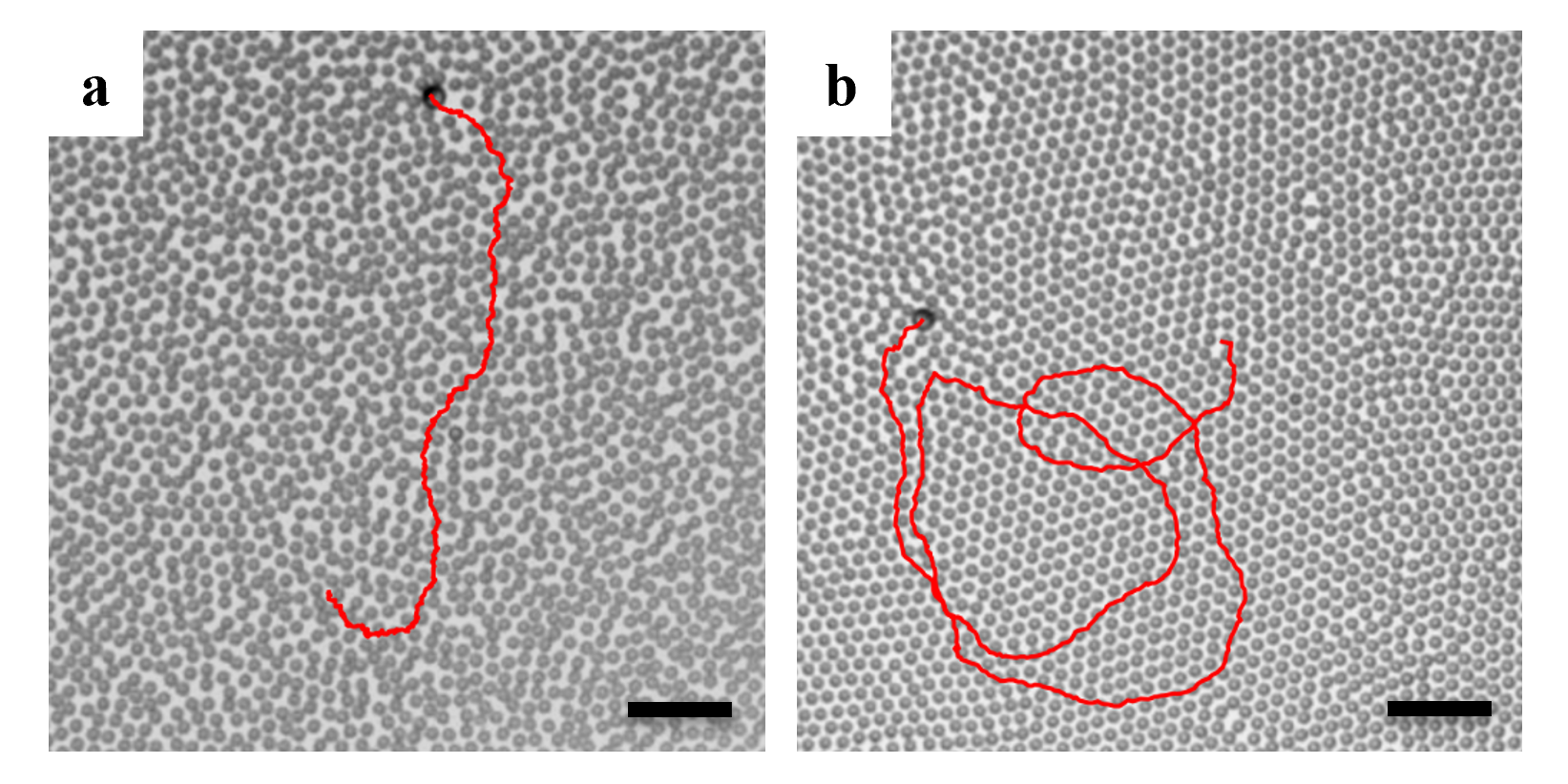}
	\caption{\textbf{Active particles in tunable colloidal environments.} Snapshots of active Janus particles cruising in colloidal monolayers of Brownian microspheres under applied electric fields (a) $\rm E=42$ $\rm V/mm$ and (b) $\rm E=108$ $\rm V/mm$ (packing fraction, $\phi=0.48$). The active particle trajectories for the last $12$ minutes are reported as a red lines. The scale bar corresponds to $20$ $\rm \mu m$ in both microscopy images.}\label{fig1}
\end{figure} 

In this work, we delve into the problem of self-propulsion in complex fluids by investigating, experimentally and theoretically, the motion of active Janus colloids (microswimmers) in a quasi two-dimensional matrix of passive Brownian microspheres (background particles) (Fig.~\ref{fig1}). Active-passive colloidal mixtures are, in fact, a useful model playground to understand how the non-equilibrium behavior of self-propelling objects affects a complex environment, and \textit{vice versa}. For example, self-propelling particles change the microstructure of a surrounding colloidal bath by injecting energy in a system which is otherwise in equilibrium. The particle's swimming force can compress local regions of the environment~\cite{kummel2015formation}, rearrange passive particles assembled in crystalline lattices~\cite{dietrich2018active}, anneal defects and grain boundaries~\cite{ramananarivo2019activity,van2016fabricating,abbaspour2024long}, deform particle networks~\cite{Szakasits2017dynamics} and alter the velocity distributions of passive components~\cite{hecht2024motility}. Conversely, the presence of passive particles boosts motility-induced phase separation~\cite{Stenhammar2015activity,dolai2018phase}, promotes group formation~\cite{dias2023environmental,jacucci2024patchy} and triggers long-range interactions between active particles~\cite{aragones2016elasticity,steimel2016emergent}. The rotational motion of self-propelling particles, and thus their persistence time, can also be drastically affected by a microstructured environment, with recent experimental studies showing enhanced rotational diffusion by up to two orders of magnitude in colloidal glasses~\cite{lozano2019active,abaurrea2020autonomously}. These results suggest a deep connection between the particle dynamics and the underlying mechanical properties of the medium. 

Since our goal is to tune, reversibly and simultaneously, both the activity of the Janus particles and the stiffness of the colloidal matrix, a careful selection and reproducible preparation of the experimental system are paramount. We opted for gold-capped Janus silica particles actuated by AC electric fields in a bath of passive silica spheres. As compared to other Janus swimmers (e.g. catalytic), AC-driven particles are particularly versatile: they do not `get stuck' in stiff environments and their propulsion direction and speed can be tuned through the frequency and magnitude of the applied electric field~\cite{boymelgreen2016propulsion,boymelgreen2018active}. Passive silica spheres also show a frequency dependent transition from attractive to repulsive interactions, each of which increases with increasing field strength~\cite{gong2001electrically}. Here, we chose the AC frequency to be $20$ $\rm kHz$, which assures (1) powerful propulsion of the active particles with the gold-capped side at their front, (2) continuously variable long-range repulsion between silica surfaces ({\sl i.e.}\ between passive spheres as well as between passive particles and the back hemispheres of the active colloids) and (3) a short-range repulsive interaction between the passive spheres and the gold cap of the Janus particles. As we increase the strength of the orthogonal AC electric field, the colloidal background freezes due to stronger repulsive interactions whereas the Janus particles become faster. Upon this quenching, the persistence time $\rm \tau$ of the active particles drops by more than one order of magnitude (compare for example the red trajectories in Fig.~\ref{fig1}(a) and Fig.~\ref{fig1}(b), or in the corresponding Supplementary Videos~S1 and S2). In particular, $\rm \tau$ transitions from being independent of the mean swimming velocity $v$ to scaling as $\tau \propto v^{-1}$. Moreover, for sufficiently large propulsion speeds and silica-silica interactions, we observe for the first time the emergence of chiral helical swimming in a bath composed of non-chiral microspheres.

The particle-resolved approach and flexible experimental system allow to identify clear ingredients leading to helical motion and reshape, reversibly and \textit{in situ}, the rotational characteristics of the active motion. The mechanism behind the faster reorientation is general and based on an interaction asymmetry of the active particle with the discrete colloidal matrix, a feature which our Janus particles share with a large number of other microswimmers. The dipolar repulsion acts in a non-central way, {\sl i.e.} the stronger interaction with the silica side naturally leads to the creation of a lever arm that generates a torque. To bring this mechanism to work, long-ranged interactions and a compressible matrix are needed. The spontaneous chiralization then occurs as a sustained symmetry breaking initially caused by an orientational fluctuation. These findings demonstrate the ability to regulate the reorientation dynamics of active particles through mechanical changes in microstructured backgrounds.

The manuscript is structured as follows. In Section~\ref{results}, we first discuss the preparation of colloidal monolayers where the inter-particle interactions can be tuned using external AC electric fields. We then describe the two-dimensional motion of active particles throughout colloidal monolayers in term of their swimming velocity and persistence time. Here, the electric field determines not only the interactions between the passive particles of the monolayer, but also the propulsion speed of the active beads. In Section~\ref{discussion}, we present a final discussion on our results. Sections~\ref{ExpSec},~\ref{NumSec} and \ref{TheoSec} contain the experimental, numerical and theoretical protocols.

\section{Results}\label{results}

\subsection{Tunable colloidal environments}

In this Section, we showcase that external AC electric fields can be used to modulate interactions among microparticles and therefore tune the structural and mechanical properties of two-dimensional colloidal environments. In this way, we are able to quench colloidal monolayers at fixed packing fraction.

\begin{figure*}[t]
  \centering
  \includegraphics[scale=0.8]{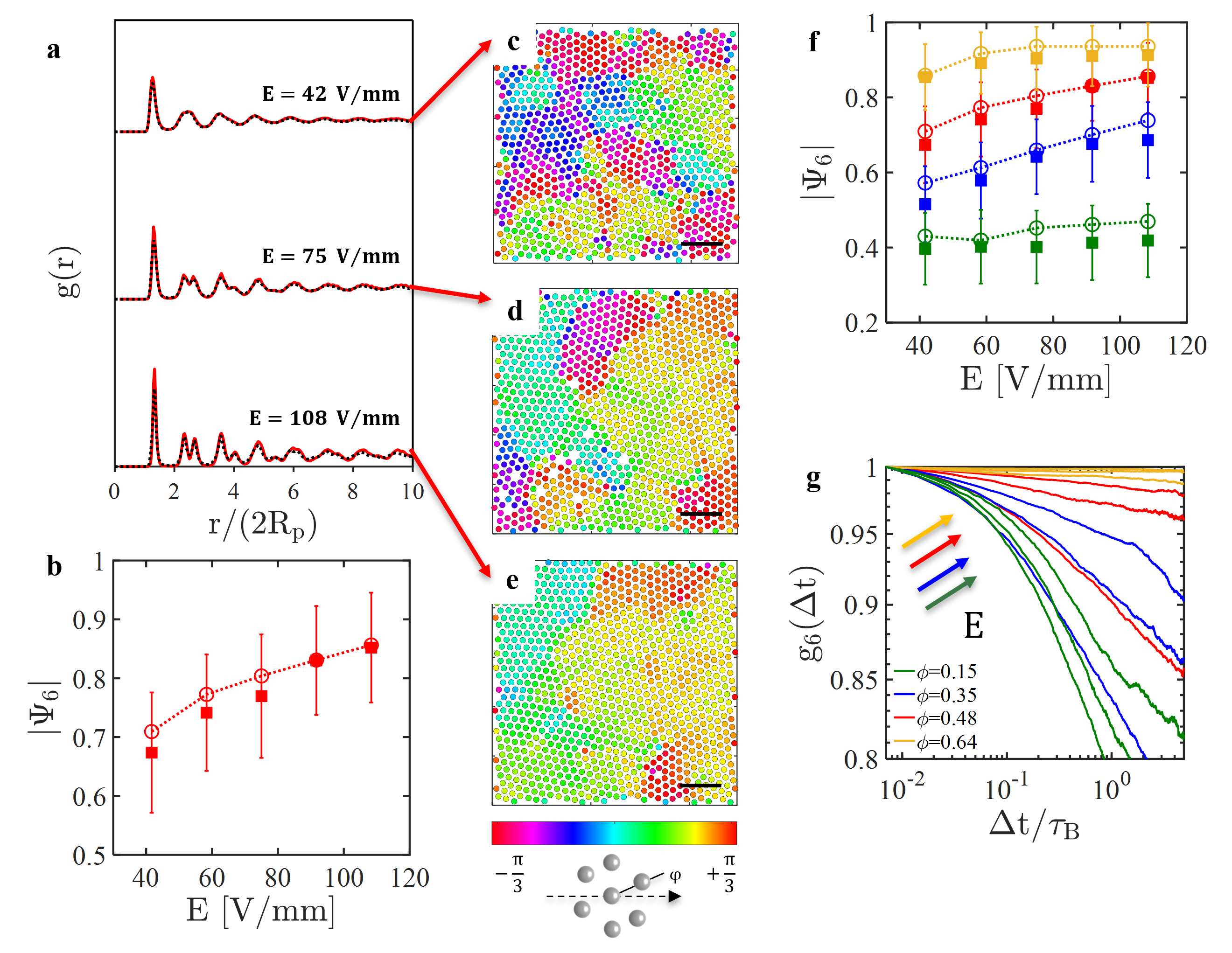}
  \caption{\textbf{Tuning the structure of the environment.}
  (a) Experimental (red solid line) and numerical (black dashed line) pair correlation functions $\rm g(r)$ at $\phi=0.48$ for different electric fields. (b) Mean hexagonal order parameter $|\Psi_6|$ versus field strength in experiments (solid symbols) and simulations (empty symbols and dashed connecting line) at $\phi=0.48$. (c–e) Experimental maps of the phase $\varphi$ of $\Psi_{6,n}$ at $\rm E=42$ $\rm V/mm$, $\rm E=75$ $\rm V/mm$, and $\rm E=108$ $\rm V/mm$. The scale bar is $20\,\mu$m and the colours denote the crystalline orientations as sketched in the bottom-right inset. (f) $\left| \Psi_{6} \right|$ plotted as a function of the electric-field strength $\rm E$ for colloidal monolayers with packing fractions $\phi=0.15$ (green), $\phi=0.35$ (blue), $\phi=0.48$ (red), and $\phi=0.64$ (orange). Filled and empty symbols correspond to experimental and numerical results, respectively. (g) Experimental dynamical orientational correlation $\rm g_6(\Delta t)$ as a function of the normalized delay time $\rm \Delta t / \tau_{B}$ for different packing fractions (according to same colour code as in (f)) and electric fields. For clarity, only the curves at $\rm E=48$\,V/mm, $\rm E=75$\,V/mm, and $\rm E=108$\,V/mm are shown. At all $\phi$, the decay of $\rm g_6(\Delta t)$ becomes slower for larger electric fields, as indicated schematically by the arrows.}
  \label{fig2}
\end{figure*}

In experiments, the colloidal environments consist of silica microspheres ($\rm SiO_2$, radius $\rm R_p = 1.46$ $\rm \mu m$) dispersed in water and sitting on a planar electrode. The packing fraction $\phi$, {\sl i.e.} the relative area occupied by the particles in the $xy$-plane, ranges from $\phi=0.15$ to $\phi=0.64$. We apply an alternating-current (AC) electric field $\rm \mathbf{E}$ in the $z$-direction at fixed frequency ($\rm f=20$ $\rm kHz$) and magnitude between $\rm E = 0$ $\rm V/mm$ and $\rm E = 108$ $\rm V/mm$. These values are chosen to assure that the spheres stay within a monolayer and are not squeezed out into a second layer. At $\rm E=0$, the colloidal particles exhibit weak short-range repulsive interactions due to the negatively charged surface of $\rm SiO_2$ and are free to laterally move close to each other as a result of Brownian diffusion. The presence of an AC electric field significantly alters the mutual interactions and overall colloidal microstructure. In particular, at $\rm f=20$ $\rm kHz$, the field polarizes the particles and induces a long-range dipole-dipole pair potential that is proportional to $\rm E^2$ and decays as $\rm r^{-3}$, $\rm r$ being the inter-particle distance~\cite{gong2001electrically,Ristenpart2003electrically}. It is the leading contribution to the inter-particle forces when the particles are sufficiently far from each other. 

In simulations, we model the background matrix as a two-dimensional system of interacting Brownian particles, where hydrodynamics interactions are negligible and the solvent simply induces a thermal noise~\cite{medina1988long}. The particles interact via volume exclusion, modeled as a WCA potential~\cite{weeks1971role}, and a long-range dipole-dipole repulsion mimicking the dipolar interactions between $\rm SiO_2$ microparticles at $\rm E > 0$. The strength of these potentials, $\epsilon$ and $\rm K$ respectively, is determined by matching the peaks of the pair correlation functions $\rm g(r)$ with experiments for every value of $\rm E$ and $\rm \phi$ (see Fig.~\ref{fig2}(a), Fig.~S1 and Table~S1). Further experimental and numerical details are reported in Sec.~\ref{ExpSec} and Sec.~\ref{NumSec}, respectively.

Both in simulations and experiments, the long-range repulsive interactions induced by the external electric field increase the effective particle size and promote a shift of the freezing line of the two-dimensional phase diagram towards lower packing fractions. Features that are characteristic of solid structures (e.g.\ hexagonal crystalline domains) appear at smaller $\phi$, when the original microstructure (at $\rm E=0$) was in the fluid phase. The red curves in Fig.~\ref{fig2}(a) show for example the pair correlation functions, $\rm g(r)$, for a monolayer at $\phi=0.48$ subjected to three different electric fields. As $\rm E$ increases, the peaks of $\rm g(r)$ become more pronounced until, at $\rm E=108$ $\rm V/mm$, the second peak between $\rm 4R_p$ and $\rm 6R_p$ splits, as reported for hexagonal structures. To quantify the appearance of solid domains, we consider the absolute value of the orientational hexagonal order parameter per particle: 

\begin{equation}\label{psi6}
    \rm \left| \Psi_{6,n} \right| = \frac{1}{N_n} \left| \sum_k e^{(i6\theta_{nk})} \right| \,,
\end{equation}

\noindent where $\rm N_n$ is the number of neighbours of the $\rm n^{th}$-particle and $\rm \theta_{nk}$ is the angle of each bond, so that $\rm \left| \Psi_{6,n} \right| = 1$ for a perfect hexagonal cell.
By averaging over all the particles, we obtain the global orientational order parameter $\rm \left| \Psi_{6} \right|$ shown in Figure~\ref{fig2}(b). The value of $\rm \left| \Psi_{6} \right|$ increases monotonically with the applied electric field, in qualitative agreement with the phase $\varphi$ of $\rm \Psi_{6,n}$ (Figs.~\ref{fig2}(c-e)) which represents the orientation of local crystalline domains, if any. As $\rm E$ increases, small ordered regions (e.g. Fig.~\ref{fig2}(c), $\rm E=42$ $\rm V/mm$) merge into large hexagonal domains (e.g. Fig.~\ref{fig2}(e), $\rm E=108$ $\rm V/mm$). The evolution is also illustrated in the Supplementary Video~S3, where the electric field is progressively increased from $\rm E=0$ to $\rm E=108$ $\rm V/mm$.

The structural changes induced by the electric field are observed for a broad range of packing fractions, suggesting the generality of the observed quenching mechanism. These results are shown in Fig.~\ref{fig2}(f), where $\rm \left| \Psi_{6} \right|$ is reported as a function of the electric field $\rm E$ for four different values of $\phi$. The corresponding pair correlation functions $\rm g(r)$ and phases $\varphi$ of $\rm \Psi_{6,n}$ are shown in Fig.~S1 and Fig.~S2, respectively. In all instances, the global bond order parameter $\rm \left| \Psi_{6} \right|$ increases with $\rm E$, although we observe only a weak increase for large values of $\rm E$ at $\phi=0.15$ (green data) and a plateau in the densest environment ($\phi=0.64$, orange data). Indeed, in the small packing fraction regime, the smaller is the density, the larger is the increase of the dipole interaction strength needed to induce significant structural changes. By contrast, at high packing fractions, the observed plateau implies that the system has already reached the almost-close packing regime characterized by a defects-poor configuration. As a further confirmation, we illustrate in Fig.~\ref{fig2}(g) the dynamical orientational correlation $\rm g_6(\Delta t)= \langle \Psi^{*}_{6,n}(\Delta t) \Psi_{6,n}(0) \rangle$ as a function of the delay time $\rm \Delta t / \tau_{B}$ normalized by Brownian diffusion time $\rm \tau_{B}=(6 \pi \eta_0 R_p^3)/(k_B T)$, where $\eta_0$ is the water viscosity and $\rm k_B T$ is the thermal energy (see also Fig.~S3 for a comparison with the numerical results). At $\phi=0.64$ (orange data), $\rm g_6(\Delta t)$ shows no significant decay within our experimental timescale, suggesting the presence of defect-free structures. At smaller packing fractions, $\rm g_6(\Delta t)$ decays at a rate that depends on both $\rm E$ and $\phi$. Interestingly, similar decays are recovered with different packing fractions and electric fields, in agreement with the measurements of $\rm \left| \Psi_{6} \right|$ (compare for instance the red and blue data in Fig.~\ref{fig2}(f)). This demonstrates that the electric field can be used as a tool to create optimally ordered configurations without changing the packing fraction.

\subsection{Active particles in tunable colloidal environments}\label{active}

\begin{figure*}[t]
  \centering
  \includegraphics[width=\textwidth]{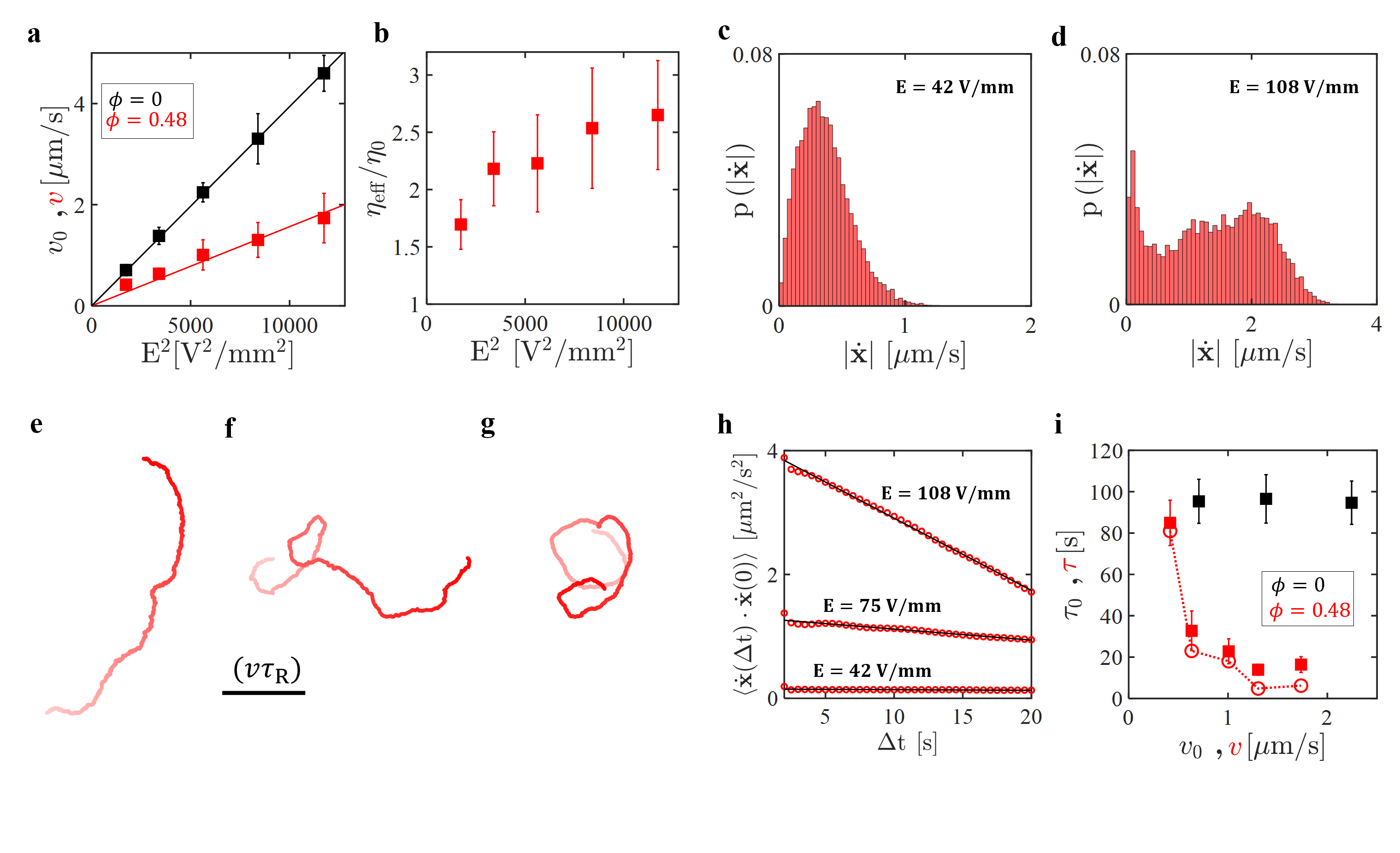}
  \caption{\textbf{Swimming in crowded environments.}
  (a) Mean swimming velocities (black) $v_0$ and (red) $v$ plotted as a function of the electric-field strength squared $\rm E^2$ for active particles swimming “freely” ($\phi=0$) or cruising in colloidal environments of packing fraction $\phi=0.48$. The solid lines are linear fits crossing the origin. (b) Ratio between the effective viscosity $\eta_{\mathrm{eff}}$ of the colloidal environment and the water viscosity $\eta_0$, calculated as described in the text.
  (c–d) Normalized histograms of the instantaneous speed $|\dot{\mathbf{x}}|$ under an applied electric field: (c) $\rm E = 42\,\mathrm{V/mm}$ and (d) $\rm E = 108\,\mathrm{V/mm}$.
  (e–g) Trajectories of active particles cruising in colloidal environments of packing fraction $\phi=0.48$ under applied electric fields: (e) $\rm E = 42\,\mathrm{V/mm}$, (f) $\rm E = 75\,\mathrm{V/mm}$, and (g) $\rm E = 108\,\mathrm{V/mm}$. The scale bar is dynamic and corresponds to $(v \tau_{\mathrm{R}})$ for all trajectories, where $v$ is the average swimming velocity and $\tau_{\mathrm{R}}$ is the rotational Brownian time (defined in the text). The color indicates the time: from $t = 0$ (white) to $t = 500\,\mathrm{s}$ (red). (h) Time autocorrelation functions of $\dot{\mathbf{x}}$ for the trajectories shown in (e–g). The solid lines are linear fits. (i) Mean persistence times $\tau$ (red symbols, $\phi=0.48$) and $\tau_0$ (black symbols, $\phi=0$) plotted against $v$. The filled symbols are obtained from experiments, while the empty symbols (linked by the dotted line) are from numerical simulations matching the experimental values of $v$. In all panels, the error bars represent the standard deviations.}
  \label{fig3}
\end{figure*}

In this Section, we demonstrate that the tunable colloidal environments dramatically alter the motion of active particles cruising through them. These changes are modulated using a single control knob: the magnitude of the AC electric field ($\rm E$). 

As model self-propelling (or active) particles, we use silica ($\rm SiO_2$) microspheres of radius $\rm R_a=2.5$ $\rm \mu m$ half-coated with $10$ $\rm nm$ of gold ($\rm Au$). They settle to the bottom substrate and, under the same AC electric fields introduced in Sec. 2.1, self-propel in the $xy$-plane with the $\rm Au$-hemisphere heading due to induced-charge electrophoresis (ICEP)~\cite{Gangwal2008induced,yan2016reconfiguring,boymelgreen2018active,van2019interrupted}. Without the colloidal environment, the magnitude of the \emph{swimming velocity} increases linearly with $\rm E^2$. Instead, the direction of motion changes according to rotational Brownian motion about the $z$-axis, happening over a timescale $\rm \tau_R = (k_B T/\xi_R)^{-1}$ ($\rm \xi_R=8 \pi \eta_0 R_a^3$ is the rotational friction coefficient). Thus, the characteristic \emph{persistence time} of the active trajectories depends neither on the swimming velocity nor on the magnitude of the external electric field.  The characterization of the `free' active motion (at $\phi=0$) is given in Fig.~S4 and typical trajectories are shown in the Supplementary Video~S4 ($\rm E=42$ $\rm V/mm$) and Supplementary Video~S5 ($\rm E=108$ $\rm V/mm$). 

We then prepare the same colloidal environments as in Fig.~\ref{fig2} and add a very small number of self-propelling particles such that the interactions between them can be neglected. The active particles behave as `snowploughs' making their way through the two-dimensional crowded environment (see Supplementary Videos~S1 and S2, note that the SiO\(_2\) spheres do not get into physical contact). Even though they do not sense local defects and grain boundaries due to the large size (compared to the one of the bath passive particles), their active dynamics is strongly coupled to the monolayer's properties (see Fig.~\ref{fig1}a and Fig.~\ref{fig1}b). In simulations, this behavior is reproduced through a 2D active Brownian particle dynamics where the active particle moves in the direction identified by the outward vector normal to the $\rm Au$ hemisphere. As for passive colloids, hydrodynamics interactions are neglected and the solvent induces translational and rotational noise on the particle position and orientation~\cite{abraham2023dynamics}. The active particle interacts with passive colloids via volume exclusion and repulsive dipolar forces. Volume exclusion is obtained from a WCA potential accounting for the larger size of the active colloid, whereas dipolar forces stem from the $\rm SiO_2$ back hemisphere of the active particle. As such, although active-passive dipolar interactions follow the $\rm ~r^{-3}$ scaling (as in the case of uncoated passive particles), the force is not applied on the particle center. Instead, it acts the center of the SiO\(_2\) hemisphere, due to the fact that the gold cap and the passive particles do not exhibit dipolar mutual interactions. Hence, the orientational dynamics is not purely governed by rotational noise but it is additionally susceptible to local torques. In the following, we analyze the active trajectories by extracting the mean swimming velocity $v$ and persistence time $\tau$. The same quantities measured at $\phi=0$ (without the colloidal environment) are denoted as $v_0$ and $\tau_0$. As in the previous Section, we first describe the main results at $\phi=0.48$ and then show the generality of our findings at other packing fractions. Further information about the particle synthesis and self-propulsion method are in Sec.~\ref{ExpSec}. The numerical simulations are described in Sec.~\ref{NumSec}. 

In experiments, we record active trajectories for any applied electric field, measure the instantaneous velocities $\dot{\mathbf{x}}$ and compute the mean swimming speed $v$ and $v_0$ as described in Sec.~\ref{ExpSec}. The latter quantities are shown in Fig.~\ref{fig3}(a) as a function of the electric-field strength squared, as red ($v$, at $\phi=0.48$) and black ($v_0$, at $\phi=0$) symbols. Even though the colloidal environment does not affect the ICEP linear relationship between $v$ and $\rm E^2$ (see solid fitting lines in Fig.~\ref{fig3}(a)), the self-propelling particles are consistently slower when immersed in a bath of passive particles ($v<v_0$ for any $\rm E$). This result can be naively rationalized by taking into account an increased effective viscosity, $\rm \eta_{eff}$, of the medium, following the addition of passive uncoated $\rm SiO_2$ beads. Moreover, $\rm \eta_{eff}$ depends on both the packing fraction $\phi$ and electric field $\rm E$, since more particles and stronger inter-particle interactions increase the effective friction produced by the environment onto the active particle. This effect is highlighted in Fig.~\ref{fig3}(b) where we plot the ratio of $\rm \eta_{eff}/\eta_0$ assuming that the swimming force $f_{\rm s}$ only depends on the applied electric field, {\sl i.e.} $f_{\rm s}=6\pi \eta_0 {\rm R_a} v_0 = 6 \pi \eta_{\rm eff} {\rm R_a} v$, being $\eta_0$ the water viscosity. This ratio approximately doubles as the external field increases from $\rm E = 42$ $\rm V/mm$ to $\rm E = 108$ $\rm V/mm$. 

Nonetheless, a description based on an average swimming velocity of the self-propelling particles and an effective viscosity of the colloidal environment does not take into account the microstructure of the monolayer. A closer look at the instantaneous velocities sometimes reveals large fluctuations around their mean value, where $\left | \dot{\mathbf{x}} \right |$ drops to nearly zero before increasing again. This local and transient caging is due to the microstructure of the surrounding monolayer as well as the strength of interaction between its building blocks~\cite{dietrich2018active,reichhardt2020directional}; it is qualitatively illustrated in the Supplementary Video~S6 and pinpointed in Fig.~\ref{fig3}(c-d) by plotting the probability distribution ${\rm p}(\left | \dot{\mathbf{x}} \right |)$. For small applied electric fields ($\rm E = 42$ $\rm V/mm$, Fig.~\ref{fig3}(c)), the dipolar interactions are weak and the active particle can navigate through the environment owing to the high mobility of the uncoated $\rm SiO_2$ microspheres. The corresponding distribution $\rm p(\left | \dot{\mathbf{x}} \right |)$ is therefore similar to those of free active particles (see also Figure~S5, gray histograms). At higher electric fields ($\rm E = 108$ $\rm V/mm$, Fig.~\ref{fig3}(d)), $\rm p(\left | \dot{\mathbf{x}} \right |)$ shows a peak at $\left | \dot{\mathbf{x}} \right | \sim 0$ followed by a distribution at higher speeds; the active motion becomes intermittent and particles are occasionally caged by the surrounding microstructure. Velocity distributions for intermediate values of $\rm E$ are reported in Fig.~S5 and show a consistent transition towards a bimodal distribution, as $\rm E$ increases. We report no caging at smaller packing fractions ($\phi=0.35$ and $\phi=0.15$). The same qualitative behaviour is also observed in numerical simulations (Fig.~S5), although caging occurs therein more often. The discrepancy is likely due to the fact that the numerical system is perfectly two dimensional, whereas in experiments the active particles protrude above the colloidal monolayer because of their larger size (see Fig.~\ref{figM1} in Sec.~\ref{ExpSec}). To support this hypothesis, Supplementary Video~S7 shows an active particle being frequently caged in a monolayer of passive spheres of similar size.  

An increase of effective viscosity due to larger $\phi$ or $\rm E$ should also imply a slowdown of rotational Brownian motion of the active particles in the $xy$-plane and, consequently, a higher persistence time of their directed motion ({\sl i.e.} a larger $\tau$). The data in Figure~\ref{fig3} suggest the opposite. Fig.~\ref{fig3}(e-g) shows three typical active trajectories at fixed $\phi=0.48$ and different applied electric fields. As $\rm E$ increases from $42$ $\rm V/mm$ (Fig.~\ref{fig3}(e)) to $108$ $\rm V/mm$ (Fig.~\ref{fig3}(g)), the particles not only become faster, but also change their swimming direction more frequently, as shown by comparing the typical straight path of each trajectories to the persistence length $(v \tau_{\rm R})$ expected as a result of free rotational diffusion with timescale $\rm \tau_R$. We extract the characteristic persistence times $\tau$ and $\tau_0$ from the linear fit (Fig.~\ref{fig3}(h), black lines) of the time autocorrelation function of the instantaneous velocity vector $\rm  \left \langle \dot{\mathbf{x}}(\Delta t) \cdot \dot{\mathbf{x}}(0) \right \rangle$ (Figure~\ref{fig3}(h), red circles), as described in Sec.~\ref{ExpSec}. Faster decays indicate smaller persistence times. At $\phi=0.48$, the velocity autocorrelation decays faster for larger applied electric fields. This corresponds to mean persistence times that become more than one order of magnitude smaller as $\rm E$ goes from $\rm E=42$ $\rm V/mm$ to $\rm E=108$ $\rm V/mm$, {\sl i.e.} as the mean swimming velocity increases (Fig.~\ref{fig3}(i), red symbols). This is in stark contrast to the motion of `free' active particles ($\phi=0$) where $\tau_0$ remains approximately constant at all applied electric fields and swimming velocities (see black symbols in Fig.~\ref{fig3}(i) as well as Fig.~S4). The measured persistence times are also in agreement with numerical simulations (Fig.~\ref{fig3}(i), empty red symbols) that are set to match the mean swimming velocity $v$ of the active particles and the pair correlation function $\rm g(r)$ of the colloidal bath. 

The mechanism of reorientation is sketched in Figure~\ref{fig4}. An active particle cruising through a colloidal environment with its $\rm Au$ hemisphere heading (Fig.~\ref{fig4}(a)) compresses the matrix ahead. However, the orientation $\mathbf{\hat{n}}$ constantly changes due to spontaneous fluctuations. Upon one sufficiently strong fluctuation, the swimming force $f_{\rm s}\mathbf{\hat{n}}= 6 \pi \eta_0 {\rm R_a}v_0\mathbf{\hat{n}}$ causes an asymmetric compression of the colloidal environment. For example, in Fig.~\ref{fig4}(b) we use springs to schematically depict a local left compression due to a counter-clockwise fluctuation by an angle $\Delta \theta$. Once left behind at the back (Fig.~\ref{fig4}(b)), the compressed region triggers a torque $\mathbf{M} = \boldsymbol{\ell} \times \Vec{F}_{\text{eff}}$, where $\boldsymbol{\ell}$ is the lever arm and $\Vec{F}_{\text{eff}}$ denotes the total dipolar force acting on the end point of the lever arm (see blue arrow). Importantly, the reorientation self-sustains as the $\rm SiO_2$-$\rm Au$ colloid moves forward; an active particle reoriented to the left by a counter-clockwise torque compresses the colloidal environment on its left-hand side (see Fig.~\ref{fig4}(b)) so that the following total dipolar force leads again to a torque in the same direction. This is also illustrated in Fig.~\ref{fig4}(c) and Fig.~\ref{fig4}(d) where we show the local packing of the environment before and after a counter-clockwise rotation. The local compression appears in the front but, as the active particle moves, it shifts to the back where it resumes the torque -- a mechanism that is based on the separation of timescales between particle motion and environment recovery. The same is true for clockwise rotations. 

According to our description, $\tau \propto v^{-1}$ since more distance travelled means more reorientation events. We verified this dependence by performing numerical simulations where $v$ (swimming velocity of the individual active particles) and $\rm K$ (strength of the dipolar interactions, see Eq.~\eqref{eq:dipdippotential} and Table~S1) are decoupled. The results are shown in Fig.~S6; in most instances, $\tau$ and $v$ are inversely proportional suggesting that each reorientation event depends weakly on the local interaction strength $\rm K$. However, since a given reorientation is amplified, the persistence time $\tau$ strongly depends on the global stiffness of the medium. The model also predicts that the magnitude of the torque acting on the Janus particle increases with $\rm E^2$ since this is the scaling of the dipolar repulsive interactions. Extracting the average torque as $\rm \langle M \rangle = \xi_{\rm R} \langle \dot{\theta} \rangle$ ($\langle \dot{\theta} \rangle$ is the mean angular speed, see also Eq.~\eqref{eq:activedynamics}b) reveals good agreement with $\rm \langle M \rangle \propto E^2$ (see black line in Fig.~\ref{fig4}(e)) in both the numerical (Fig.~\ref{fig4}(e), empty circles) and experimental data (Fig.~\ref{fig4}(e), solid squares), proving that that interaction asymmetry and long-range repulsions are key ingredients to achieve self-sustaining reorientations. In a matrix of passive particles interacting only via WCA forces and in the absence of caging effects $\tau$ remains approximately constant (see Fig.~S7). Likewise, no self-sustaining reorientations occur if we preserve the long-ranged dipolar interactions, but set $\boldsymbol{\ell}=0$, since the orientational and translation dynamics are decoupled.

\subsection{Spontaneous chiralization}\label{chiral}

\begin{figure*}[t]
  \centering
  \includegraphics[width=\textwidth]{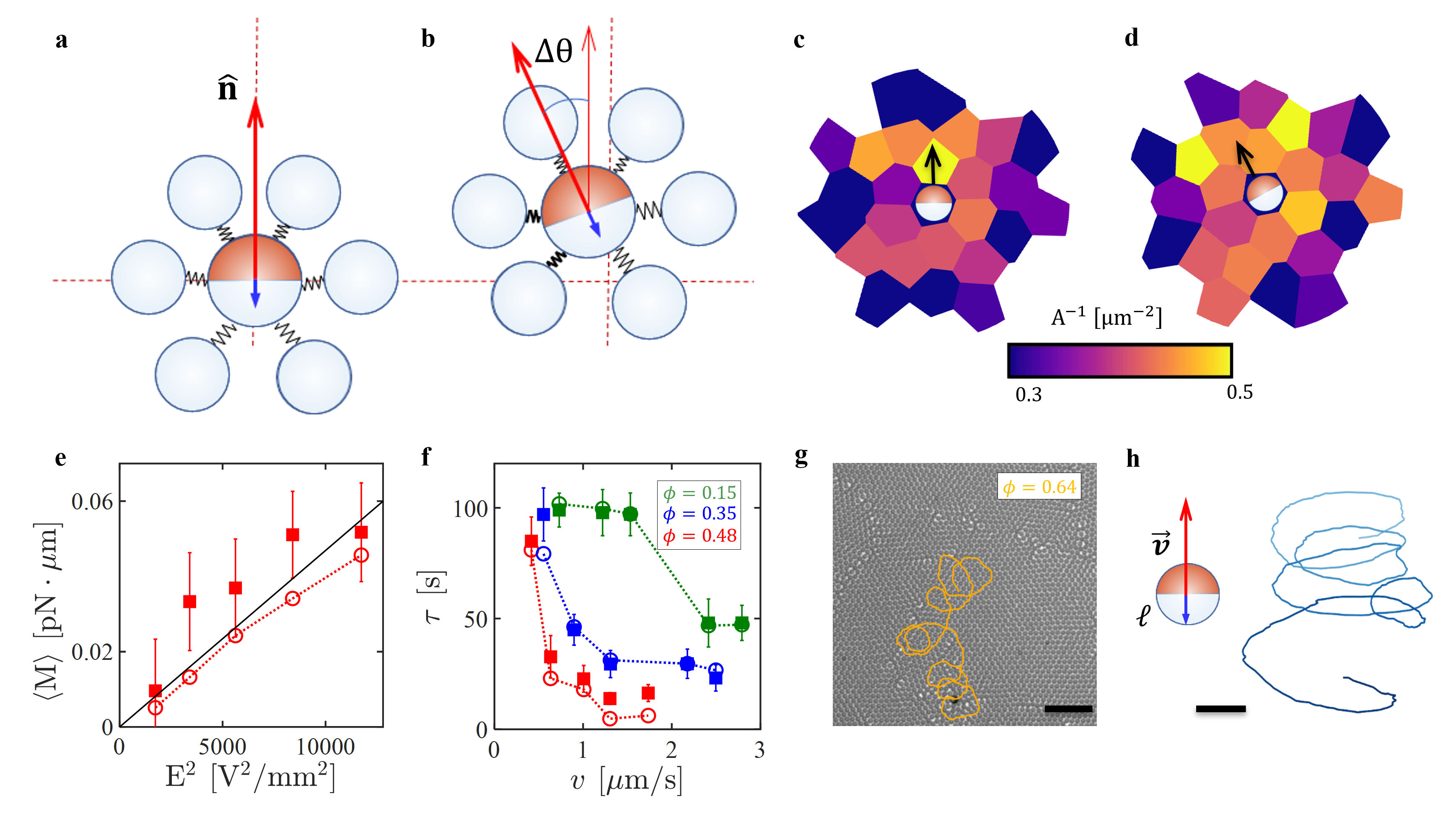}
  \caption{\textbf{Self-sustaining reorientation.}
  (a–b) Sketch of the reorientation mechanism, as modeled in numerical simulations, causing the drop of the persistence time of the active particle in colloidal environments. (a) An active particle moving with the Au cap heading produces a swimming force $f_{\rm s}\hat{\mathbf{n}}$ in the same direction. (b) Upon rotation by an angle $\Delta\theta$, the colloidal environment is compressed, generating a torque due to dipolar repulsion between passive particles and the silica back hemisphere of the active colloid. (c–d) Voronoi tessellation of the environment (c) before and (d) after a rotation. The color code indicates the local packing, i.e., the inverse area of the Voronoi cell. (e) Mean torque experienced by an active Janus particle swimming throughout a monolayer at area fraction $\phi=0.48$ for different applied electric fields. Solid and empty symbols correspond to experiments and simulations, respectively. The red dashed line connects the numerical data, and the black solid line marks the linear relationship between $\rm \langle M \rangle$ and $\rm E^2$. (f) Persistence time $\tau$ as a function of the average velocity $v$ in colloidal environments at different $\phi$, as indicated in the legend. The filled symbols correspond to the experiments, while the empty symbols are from numerical simulations matching the experimental values of $v$. (g) Helical active trajectory of $4$-minutes duration observed in experiments when the area fraction of the surrounding colloidal bath is $\phi=0.64$. The scale bar corresponds to $20\,\mu\mathrm{m}$. (h) Numerical trajectory of an active particle swimming at velocity $v=2.4\,\mu\mathrm{m/s}$ in colloidal environments at $\phi=0.35$. The color indicates the time, from $t=0$ (light blue) to $t=40$\,s (blue). The lever arm $\ell=0.4\,\mu\mathrm{m}$ determining the torque is shown as a blue arrow pointing from the geometric particle center to the interaction center. The scale bar corresponds to $1\,\mu\mathrm{m}$.}
  \label{fig4}
\end{figure*}

The scenario depicted so far also occurs over a broad range of packing fractions $\phi$. Figure~\ref{fig4}(f) summarizes the dependence of $\tau$ on $v$ in colloidal environments at different packing fraction: $\rm \phi=0.15$ (green data), $\rm \phi=0.35$ (blue data) and $\rm \phi=0.48$ (red data, redrawn from Fig.~\ref{fig3}(i)). In all instances, we report a drop of persistence time as the swimming velocity increases. Only for small $v$ and $\phi$ (first three green data points) or large $v$ and $\phi$ (last two red data points), $\tau$ remains roughly constant. The observed behavior is reproduced by numerical simulations (empty symbols and connecting dotted lines) remarkably well. Note, however, that the simple model described in Fig.~\ref{fig4}(a-d) breaks down if the environment is too sparse (at $\phi=0.15$ for small $\rm K$ values) and for strong velocity fluctuations, {\sl i.e.} upon caging (at $\phi=0.48$ and $\phi=0.64$, for large $\rm K$ values). Finally, in experiments at $\phi = 0.64$ a significant number of active particles undergo helical motion (Fig.~\ref{fig4}(g) and Supplementary Video~S8). They behave as if they were chiral, with orbit's radii as large as $\sim 5 \rm R_p$, but no preferential kick-off direction. Because both the translational and angular velocity grow linearly with $\rm E^2$ (Fig.~\ref{fig3}(a) and Fig.~\ref{fig4}(e)), the radius of the orbit, $v/\langle \dot{\theta} \rangle$, does not depend on $\rm E$. The same break of polar symmetry is observed in numerical simulations but, at $\phi=0.64$, the spontaneous chiralization is hindered by strong caging effects which, as discussed above, are more pronounced in simulations than experiments. To recover the same phenomenology reported in experiments, we keep the packing fraction relatively low ($\phi=0.35$) and virtually shift the point on which dipolar interactions act from the center of the $\rm SiO_2$ hemisphere towards the edge (see blue arrows in the sketch of Fig.~\ref{fig4}(h)). Practically, this operation increases the lever arm $\ell$ and amplifies the torque exerted by the passive particles due to dipolar interactions, while keeping the swimming velocity and stiffness of the medium fixed. As the torque increases, the active colloid is bound to make a helical motion -- a signature of chirality. As in experiments, there is no preferential direction of rotation.

The spontaneous chiralization can be also explained through a coarse-grained theory derived from the stochastic dynamics describing the system, namely the active colloid and passive particles of the environment. Starting from the Fokker-Planck equation for the $\rm N$-body problem and integrating over all the particles of the environment, we derive an effective Boltzmann equation for the single-body probability distribution $f(\mathbf{x}, \hat{\mathbf{n}}, \rm t)$ to observe an active particle at time $\rm t$ with position $\mathbf{x}$ and orientation $\hat{\mathbf{n}}$:
\begin{equation}\label{eq:Boltzmannmain}
	\partial_{\rm t} f = \nabla \cdot \left( {\rm D_{eff}}\nabla  -  v(\rho)\hat{\mathbf{n}}\right) f  + {\rm D_{R}} \mathcal{L}_{\rm a} f  - \frac{\langle \dot{\theta} \rangle}{\xi_{\rm R}}\frac{\partial}{\partial \hat{\mathbf{n}}}\cdot \left[\mathbf{z}\times \hat{\mathbf{n}}\right]f \,.
\end{equation}
where $\mathbf{z}$ is a unit vector normal to the plane of motion and $\mathcal{L}_{\rm a} =\frac{\partial }{\partial \hat{\mathbf{n}}} \cdot\Bigl(  \hat{\mathbf{n}} +  \frac{\partial }{\partial \hat{\mathbf{n}}}\cdot\boldsymbol{\mathcal{D}}\Bigl)$ is generated by the angular part of the dynamics, as in previous studies~\cite{marconi2021hydrodynamics} (see Sec.~\ref{TheoSec} for the definition of $\mathcal{D}$ and the derivation of Eq.~\eqref{eq:Boltzmannmain}). The interactions with the passive particles of the environment induce three main effects: (i) they generate a density-dependent swimming velocity which is $v(\rho)=v_0 - \rho \zeta$, where $\zeta$ is a constant term  whose expression is provided in Sec.~\ref{TheoSec}; (ii) they induce an effective diffusion coefficient $\rm D_{\text{eff}} < D_a$ (with $\rm D_a = (k_B T)/(6 \pi \eta R_a)$); (iii) they give rise to an effective angular drift velocity, $\langle \dot{\theta} \rangle$, or chirality, which explains the helical trajectories observed experimentally and numerically. While (i) and (ii) are expected from previous results~\cite{speck2015dynamical}, (iii) is a new term. Since the dipolar force is not applied to the particle center of mass but to the center of the $\rm SiO_2$ hemisphere, the angular dynamics is governed by a net torque arising from gradients in the density of the environment:

\begin{equation}
	\boldsymbol{\mathcal{M}} \approx \Delta \rho \pi \lambda^2 \,{\rm \tilde{g}}\, |\tilde{\mathbf{F}}^{\rm dip} | \ell \,\hat{\mathbf{z}}
\end{equation}
Here, $\Delta \rho$ denotes the density change and $\lambda$ corresponds to the typical length governing the dipolar interactions. The term $\rm \tilde{g}$ is the pair correlation function while $|\tilde{\mathbf{F}}^{\rm dip}|$ is the dipolar force evaluated at the contact point between an active and a passive particle. The density change $\Delta \rho$ can be expressed as the volume change due to the self-propulsion force. Equivalently, the pressure change $\Delta \rm p$  corresponds to the swim pressure due to the activity:
\begin{equation}
	\label{eq:app_eq_Deltarho}
	\Delta \rho = -\rho_0 \frac{\Delta \rm V}{\rm V} = \rho_0 \chi \Delta {\rm p} =\rho_0 \chi \frac{\gamma v_0}{2\pi \lambda}
\end{equation}
where $\chi$ is the compressibility of the passive environment. The angular velocity $\langle \dot{\theta} \rangle$ is then approximated as
\begin{equation}
	\label{eq:prediction}
	\langle \dot{\theta} \rangle= \frac{|\boldsymbol{\mathcal{M}}|}{\xi_{\rm R}}=\rho_0 \chi \frac{\gamma v_0}{2 \xi_{\rm R}}  \lambda \,{\rm \tilde{g}}\, |\tilde{\mathbf{F}}^{\rm dip} | \ell \,.
\end{equation}
In agreement with the experimental and numerical results, $\langle \dot{\theta} \rangle$ increases linearly with the self-propulsion speed $v_0$, and thus vanishes in equilibrium or for a force that is applied to the geometric center of the particle ($\ell =0$). In addition, this term disappears when the local density change of the environment around the active particle is negligible, as in experiments with a low-density background, or when the background is incompressible, e.g. in a colloidal glass made of hard spheres. The full coarse-grained theory is derived and reported in Sec.~\ref{TheoSec}.

\section{Discussion and Conclusions}\label{discussion}   

After presenting an experimental strategy to tune the structure of colloidal monolayers at fixed packing fraction, we investigated the dynamical properties of active colloids cruising throughout them. The active motion is strongly affected by the environment, which alters the particle speed and reorientation, and even gives rise to unexpected helical motion (a qualitative summary of the observed types of motion is in Fig.~S8). The mechanical properties of the environment and microswimmers activity are adjusted reversibly and \textit{in situ} using the same experimental control knob: the magnitude of an applied AC electric field.

The role of orthogonal AC electric fields in determining the pair interaction between charged microparticles sitting onto planar electrodes has been extensively investigated in the past decades~\cite{Edwards2014controlling,Bazant2004induced,nadal2002electrically,Kusner1994twostage}. In particular, the AC frequency is paramount since it regulates whether the interactions between particle pairs are attractive~\cite{Mittal2008polarization,Tanaku2023island,ristenpart2004assembly,lumsdon2004twodimensional} or repulsive~\cite{Kusner1994twostage,Ristenpart2003electrically,gong2017electricfield}. At our operating frequency $\rm f = 20$ $\rm kHz$, long-range dipole-dipole repulsive interactions between $\rm SiO_2$ particles are dominant; their strength ($\rm K$, Eq.~\eqref{eq:dipdippotential}) is always much larger than the energy scale of the WCA potential ($\epsilon$, Eq.~\eqref{eq:WCApotential}) at the typical experimental inter-particle distances identified by the first peak of the pair correlation function (Fig.~\ref{fig2}(a), Fig.~S1 and Table~S1). As such, the experimental system presented in Fig.~\ref{fig2} is similar to the colloidal monolayers described in Refs.~\cite{Kusner1994twostage,Deutschlander2013twodimensional,deutschlander2015kibble}, where quenching (or melting) took place under external electric ($\rm f > 1$ $\rm MHz$) and magnetic fields. These works reported phase transitions (from liquid to hexatic and hexatic to crystal) occurring at well-defined combinations of packing fraction and external-field magnitude. The detailed characterization of the colloidal phases, which was also performed in quasi-two-dimensional experiments of colloidal `hard spheres'~\cite{Thorneywork2017two}, calls for the determination of the exact functional decay of the dynamical orientational correlations shown in Fig.~\ref{fig2}(g). These measurements would required times or length scales that are much larger than what we can currently achieve with our experimental setup. Here, our focus is on the influence of colloidal background, as it becomes more quenched, on the active motion of self-propelling particles. Nonetheless, the curves in Fig.~\ref{fig2}(g) are in qualitative agreement with the existing literature and highlight the possibility of freezing (or melting) colloidal monolayers at fixed $\phi$ using external $\rm kHz$ AC electric fields. 

We then demonstrated that the tunable colloidal environments dramatically alter the trajectories of active colloids actuated by the same electric fields. In particular, during the quenching of the surrounding monolayers, the persistence time $\tau$ of the active Janus particles becomes dependent on their swimming velocity $v$. It rapidly drops by more than one order of magnitude as $v$ increases. A similar decrease of persistence time (often denoted as `rotational diffusion enhancement') has been reported for self-propelling Janus particles in polymer solutions ~\cite{narinder2018memory,saito2025self,gomez2016dynamics} and initially ascribed to viscoelasticity and memory effects. The same experimental results were recently revised by De Corato \textit{et al.} using a memory-free model based on local density variations, anisotropic interactions and advection terms~\cite{decorato2021spontaneous,decorato2025enhanced}. The first two ingredients are also present in our ``dog chasing its tail'', but the advection of background particles is absent since hydrodynamic interactions are neglected. Local density fluctuations are here rather due to local compressions caused by the swimming force of the Janus particle. They occur regardless of the phase, ordering, or packing fraction of the colloidal medium, provided that the monolayer is sufficiently dense and caging effects are minimal. Under these conditions, a simple relationship between swimming velocity $v$ and persistence time $\tau$ is reported: $\tau \propto v^{-1}$. Above a critical velocity, the polar symmetry of the particles is broken and the active particle undergoes helical motion, {\sl i.e.} it behaves as a chiral swimmer. This chiralization is very different from the circular motion reported for L-shaped particles~\cite{kummel2013circular} and modular swimmers~\cite{niu2018modular,madden2022hydrodynamically,ni2017hybrid,feldmann2019light} because neither the active colloids nor the bath particles are asymmetric or bind to each other. Instead, orbiting is sustained by local compressed regions created by the active particle itself.  

In summary, our findings demonstrate the ability to tune \textit{in situ} the reorientation dynamics of active particles, which may have significant implications for medical applications ranging from micro-surgery to targeted drug delivery. Regardless the details of our experimental system, the microscopic mechanism of reorientation is general as it only requires a compressible medium and anisotropic interactions. We therefore envisage that, beyond regulating active-particle motion, our results will shed light on how compressible micro- or nanostructured fluids affect the swimming behavior of microorganisms typically comprising a `head' and `tail'. For example, catalytic micromotors, which are Janus just like our active particles, may exhibit a similar motion  when added to muddy waters for wastewater treatments or environmental monitoring. Similarly, the body and flagella of bacteria may absorb/repel macrosolutes (e.g. polymers) in a different manner leading to local compressed/depleted regions in the environmental fluid.

\section{Experimental Section}\label{ExpSec}

\subsection{Synthesis of $\rm SiO_2$-$\rm Au$ particles}

$\rm SiO_2$-$\rm Au$ particles were prepared using the drop-casting method as originally described by Yan et al.~\cite{yan2016reconfiguring}. In a pre-synthesis cleaning procedure, the surface of a glass slide was treated with piranha solution (concentration $\rm H_2SO_4:H_2O_2$: 1:3) for 3-4 hours, then sonicated and rinsed with deionized (DI) water. A dilute suspension of silica particles in DI water ($\rm SiO_2$, radius $\rm R_a=2.5$ $\rm \mu m$, Bangs Laboratories, USA) was spread over the glass slide. The slide was then tilted to allow the suspension to flow down the slide leaving a monolayer of particles. After drying, three layers are subsequently deposited onto the exposed hemisphere of the particles: approx. $3$ $\rm nm$ of chromium, $15$ $\rm nm$ of gold and $15$ $\rm nm$ of silica. The additional silica layer prevents `sticking' of the bath particles onto the gold hemisphere of the Janus colloids. The microparticles were finally detached from the glass surface by sonicating the slide for a few seconds in DI water.

\subsection{Sample preparation}

\begin{figure}[h!]
	\includegraphics[scale=0.5]{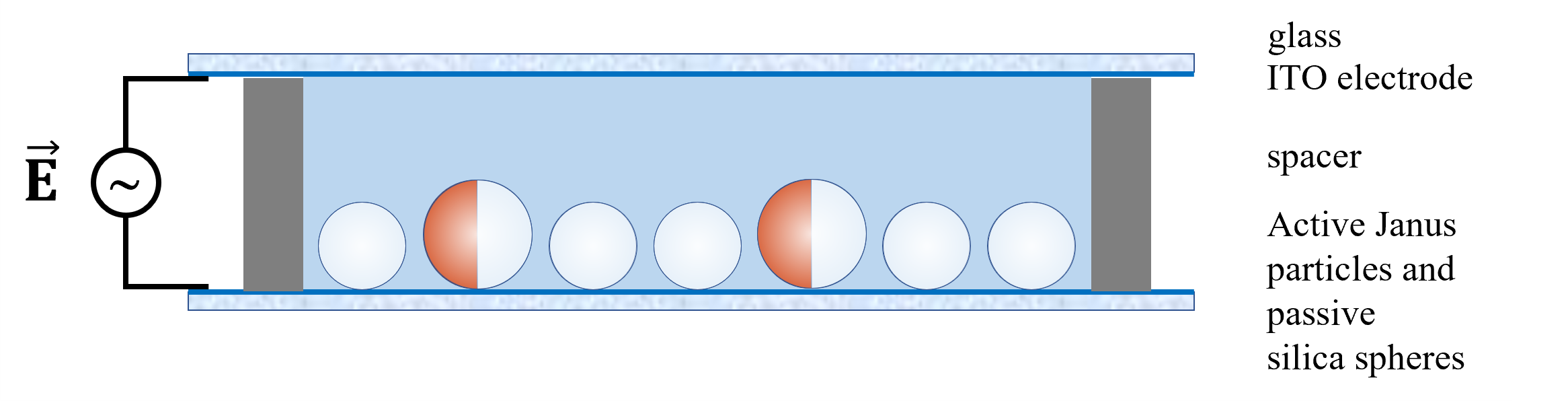}
	\caption{\textbf{Sketch of the experimental setup}. Coated and uncoated silica particles self-assemble in the $xy$-plane just above the bottom electrode and AC electric fields are applied in $z$.}\label{figM1}
\end{figure}

In each experiment, two indium tin oxide (ITO) coated glasses (surface resistivity 25–30 $\Omega \rm sq^{-1}$, Solems S. A) were first sonicated for 15 minutes in a 2\% Hellma solution, followed by cleaning in DI water for 15 minutes and blowing with $\rm N_2$. An aqueous suspension of few $\rm SiO_2$-$\rm Au$ particles and several untreated $\rm SiO_2$ particles (radius $\rm R_p=1.41$ $\rm \mu m$, Microparticles GmbH, Berlin, Germany) is then pipetted onto the slide. The cell is sealed using the second ITO slide and a double-sided adhesive circular spacer of thickness $120$ $\rm \mu m$ (Grace Bio-Labs secure seal). We let the particles settle onto the bottom surface and allow the system to equilibrate for approximately 30 minutes. The amount of $\rm SiO_2$-$\rm Au$ is always very small, such that there are no more than $5$ particles in a field of view of $800 \times 500$ $\rm \mu m^2$. Instead, the packing fraction $\phi$ is defined as the relative area occupied by the uncoated $\rm SiO_2$ microspheres in the $xy$-plane; it ranges from $\phi = 0$ (experiments of freely self-propelling particles) to $\phi=0.64$.

The two-dimensional mixtures are subjected to AC electric fields of frequency $\rm f = 20$ $\rm kHz$ and magnitude up to $\rm E = 108$ $\rm V/mm$ in the direction normal to the conductive substrate (see sketch in Fig.~\ref{figM1}). The applied electric field $\rm E$ polarizes the $\rm SiO_2$ particles and induces long-range repulsive pair interactions whose strength increases with $\rm E$. In the case of $\rm SiO_2$-$\rm Au$ colloids, the metallic and dielectric hemispheres are also polarized differently, which leads to local slip flows and two-dimensional active motion of the particles with the gold cap facing forward. This mechanism is known as induced charge electrophoresis (ICEP) and the swimming velocity is controlled by electric-field magnitude $\rm E$. The electric field also orients the $\rm SiO_2$-$\rm Au$ particles such that the axis linking the poles of the two hemispheres is parallel to the underlying substrate, which ensures that the active motion is two dimensional at all times. Importantly, the operating frequency $\rm f = 20$ $\rm kHz$ is a key parameter to observe both long-range repulsion between uncoated $\rm SiO_2$ particles and ICEP self-propulsion of $\rm SiO_2$-$\rm Au$ colloids. For example, at $\rm f \sim 1$ $\rm kHz$, the electric field drives strong solvent electrodynamic flows (EHD) along the conductive substrates in the direction of the silica microspheres~\cite{Bazant2004induced}. This effect changes the interactions between uncoated $\rm SiO_2$ spheres from repulsive to attractive and swaps the swimming direction of the $\rm SiO_2$-$\rm Au$ colloids~\cite{Tanaku2023island}. 

\subsection{Imaging and data analysis}

The sample cell is placed onto the stage of an inverted microscope (Leica DMI3000 B, Leica Microsystems GmbH, Wetzlar, Germany), where imaging is performed with $10 \times$ and $20 \times$ magnification in bright-field mode. The microscope is coupled to a CMOS camera (Basler ACE) of resolution 1280 $\times$ 1080 pixels and videos are recorded at 2 frames per second. The centres of the particles are detected over time using python tracking codes (http://soft-matter.github.io/trackpy/v0.5.0/). We record approximately $15$ active trajectories for any applied electric field and packing fraction, and calculate the instantaneous velocity $\dot{\textbf{x}}$ of each particle as the distance travelled in the time interval $\rm \Delta t = 2$ $\rm s$. From the magnitude and direction of $\dot{\textbf{x}}$ we compute the mean swimming velocity $v$ (or $v_0$ at $\phi=0$) and persistence time $\tau$ (or $\tau_0$ at $\phi=0$). Specifically, to calculate $v$ and $v_0$, we first average the instantaneous velocities of each active particle $i$ to obtain its mean swimming velocity $v_i$ and then take the mean among all the $\sim 15$ particles. To calculate $\tau$ and $\tau_0$, we consider the time autocorrelation function of the instantaneous velocity vector $\rm  \left \langle \dot{\mathbf{x}}_i(\Delta t) \cdot \dot{\mathbf{x}}_i(0) \right \rangle$ of the active particle $i$~\cite{wang2015enhanced}, which decays as

\begin{equation}\label{Autocorrelation}
	\left \langle \dot{\mathbf{x}}_i(\Delta {\rm t}) \cdot \dot{\mathbf{x}}_i(0) \right \rangle \sim v_i^2 \left (1- \frac{\Delta {\rm t}}{\tau_i} \right )\,,
\end{equation}  

\noindent where $v_i$ is the swimming velocity of particle $i$ and $\rm \Delta t$ is the delay time. Equation~\eqref{Autocorrelation} is valid for $\rm \Delta t \lesssim \tau_i$~\cite{wang2015enhanced}. $\tau$ and $\tau_0$ are the mean value of $\tau_i$ among $\sim 15$ particles. Note that, at $\phi=0$, the velocity-autocorrelation method yields values of $\tau$ that are similar to those measured from the fit of the translational mean squared displacement of the self-propelling particles (see Fig.~S4), as commonly done in several active-colloids studies~\cite{bechinger2016active}.

\section{Numerical Section}\label{NumSec}

Numerical simulations are performed by considering $\rm N$ Brownian particles with dipole-dipole interactions to model $\rm SiO_2$ colloids in solution. The self-propelling $\rm SiO_2$-$\rm Au$ colloid is modeled using active Brownian dynamics and mainly interacts with the passive environment through the dipole-dipole forces generated by the $\rm SiO_2$ back hemisphere.

\subsection{Passive particles dynamics} 

In the absence of the SiO\(_2\)-Au colloids, the passive environment consists of $\rm N$ interacting particles evolving with passive Brownian dynamics for the particle position $\mathbf{x}_i$, given by
\begin{equation}\label{eq:passive_dynamics}
    \xi_p\dot{\mathbf{x}}_i = \mathbf{F}^{\rm p-p}_i + \xi\sqrt{\rm 2D_p} \, \boldsymbol{\eta}_i \,,
\end{equation}
where $\boldsymbol{\eta}_i$ is a Gaussian white noise vector with zero average and unit variance. The terms $\xi_p$ and $\rm D_p$ are the drag friction and the diffusion coefficient of the passive particles. They satisfy the Einstein relation with the environmental temperature, $\rm D_p \xi_p = k_B T$, where $\rm k_B$ is the Boltzmann constant. Two passive particles interact via pure repulsive forces $\mathbf{F}^{\rm p-p}_i$ which are determined by two contributions: short-range volume exclusion forces and long-range dipole-dipole repulsive interactions. Both interactions are conservative and can be derived from a pairwise potential, $\mathbf{F}^{\rm p-p}_i = -\nabla_{\mathbf{x}_i} \sum_{j<k} {\rm U_{tot}}(x_{jk})$, where $x_{jk}=|\mathbf{x}_j-\mathbf{x}_k|$ is the distance between the $j$-th and $k$-th particles. The total potential $\rm U_{tot}$ is the sum of a Weeks-Chandler-Andersen (WCA) potential, $\rm U_{WCA}(r)$, and a repulsive dipole-dipole potential, $\rm U_{dip}(r)$, where $r$ is the inter-particle distance. The Weeks-Chandler-Andersen (WCA) potential has the form
\begin{equation}
	\rm U_{WCA}(r) =
		4\epsilon \left[ \left( \frac{\sigma}{r} \right)^{12} - \left( \frac{\sigma}{r} \right)^{6} + \frac{1}{4} \right]
  \label{eq:WCApotential}
\end{equation}
for $\rm r < 2^{1/6}\sigma$ and $0$ otherwise, while the dipole-dipole potential reads
\begin{equation}
    \rm U_{dip}(r)=K \left(\frac{\sigma}{r} \right)^3 \,,
    \label{eq:dipdippotential}
\end{equation}
for $\rm r<5\sigma$ and zero otherwise. In both potentials, $\sigma$ represents the distance between the centers of two interacting particles which, for interactions between passive particles, is given by the diameter of a $\rm SiO_2$ particle, $\rm \sigma= 2R_p$. $\epsilon$ and $\rm K$ are the energy scales of the two potentials.

\subsection{Active particle dynamics} 

The SiO\(_2\)-Au colloid with position $\mathbf{x}$ is modeled as an active Brownian particle. This particle moves at constant velocity along the unit vector $\hat{\mathbf{n}}=(\cos\theta, \sin\theta)$, where $\theta$ is the orientational angle determined by the normal direction of the Au hemisphere.  
Specifically, the active particle dynamics for $\mathbf{x}$ and $\theta$ read
\begin{subequations}
\label{eq:activedynamics}
\begin{align}
	&\xi_a \dot{\mathbf{x}} = f_{\rm s} \mathbf{\hat{n}} + \mathbf{F}^{\rm a-p} + \xi_a\sqrt{\rm 2D_a} \, \boldsymbol{\eta} \\
    &\xi_{\rm R}\dot{\theta} = \xi_{\rm R}\sqrt{\rm 2D_R} w + \rm M \,,
\end{align}
\end{subequations}
where $\boldsymbol{\eta}$ and $w$ are two Gaussian white noises with zero average and unit variance. $\rm D_a$ and $\xi_a$ denote the translational diffusion and the translational friction coefficient of the active particle satisfying the Einstein's relation with the environmental temperature. The term $f_{\rm s}$ denotes the self-propelled force which generates the swim velocity, $\rm D_R$ represents the rotational diffusion coefficient, $\rm \xi_R$ corresponds to the rotational friction coefficient and $\rm M$ is any torque acting the particle.

When the SiO\(_2\)-Au active particle moves in the passive colloidal monolayer, additional forces emerge because of the interactions between active and passive particles. Specifically, the dynamics in Eq.~\eqref{eq:passive_dynamics} for the $i$-th passive particle is subject to interactions with the active particle $\mathbf{F}^{\rm a-p}_i$. Reciprocally, the dynamics of the SiO\(_2\)-Au particle in Eq.~\eqref{eq:activedynamics} is governed by the interactions with the surrounding passive colloids, such that $\mathbf{F}^{\rm a-p}_a = -\sum_j \mathbf{F}^{\rm a-p}_j$, where $j=1, ..., {\rm N}$ runs over the $\rm N$ passive particles. The force $\mathbf{F}^{\rm a-p}_j$ is determined by two contributions: i) a WCA potential to model volume exclusion between passive and active particles and ii) a dipolar potential generated by the interactions between the $\rm SiO_2$ passive colloid and the $\rm SiO_2$ side of the SiO\(_2\)-Au active particle, {\sl i.e.} $\mathbf{F}^{\rm a-p}_j= - \nabla_{\mathbf{x}_j} \left(\rm  U_{WCA} + U_{dip}^a \right)$. As in the passive case, $\rm U_{WCA}$ is a pure repulsive WCA potential and has the form shown in Eq.~\eqref{eq:WCApotential} with $\sigma =\rm R_p+R_a$ representing the distance between the centers of an active and a passive particle. By contrast, the energy scale $\epsilon$ is maintained at the same value of passive-passive interactions.

The potential $\rm U_{dip}^a$ has the same functional form of $\rm U_{dip}$ (Eq.~\eqref{eq:dipdippotential}). However, since this force is generated by the SiO\(_2\) hemisphere of the active colloid, this potential differs from $\rm U_{dip}$. In the expression for $\rm U_{dip}^a(r)$, the distance $\rm r$ is the distance from the center of SiO\(_2\) hemisphere to the center of a neighboring passive particle rather than the distance calculated from the center of the active colloid. Therefore, such a dipolar force exerts a larger repulsion on the passive particles close to the $\rm SiO_2$ hemisphere compared to the repulsion exerted on the particles close to the $\rm Au$ hemisphere. Due to the selective repulsion on the SiO\(_2\) side, a lever arm of length $\ell$ arises. Consequently, the active particle is subject to a torque $\Vec{M}=\rm M\hat{\vec{z}}$ which is 
is given by:
\begin{equation}
	\Vec{M} = \boldsymbol{\ell} \times \Vec{F}_{\text{eff}} \,,
\end{equation}
where $\boldsymbol{\ell}$ is the vector which links the center of the particle with the center of the SiO\(_2\) hemisphere and $\Vec{F}_{\text{eff}}$ is the total force acting on the center of the SiO\(_2\) hemisphere arising from all passive particles. The torque $\Vec{M}$ governs the orientational dynamics of the active particle (Eq.~\eqref{eq:activedynamics}).

\subsection{Simulation details}

To identify structural changes in passive environments due to the electric field, we perform simulations in two dimensions with the dynamics of Eq.~\eqref{eq:passive_dynamics}, i.e.\ without the active particle. By contrast, the results of Sec.~\ref{active} are obtained by simulating active dynamics of Eq.~\eqref{eq:activedynamics} in a passive colloidal monolayer. With and without the active tracer, numerical simulations are performed in a box of size $\rm L$ with periodic boundary conditions. As in experiments, we consider passive environments with area fractions $\rm \phi=N \pi R_p^2/L^2 =0.15, 0.35, 0.48, 0.64$, with $\rm N=10^3$ and $\rm L$ adjusted correspondingly. Length and time are rescaled by the passive particle diameter $\rm 2 R_p$ and the WCA energy scale $\rm R_p^2 \xi/\epsilon$. With this choice, the system is governed by the following dimensionless parameters: i) the ratio between active and passive particle diameter $\rm R_a/R_p$; ii) the reduced translational diffusion coefficient $\rm \xi_p D_p/\epsilon$; iii) the reduced rotational diffusion coefficient $\rm D_R \xi_p R_p^2/\epsilon$; iv) the reduced dipolar interaction strength $\rm K/\epsilon$; v) the reduced strength of the self-propelled force $f_{\rm s} {\rm R_{p}} /\epsilon$. Finally, we point out that the torque does not generate additional dimensionless parameters being determined by i), iv), and by the particle geometry, e.g.\ the center of the $\rm SiO_2$ hemisphere.

\begin{figure*}[t]
	\centering
	\includegraphics[scale=0.8]{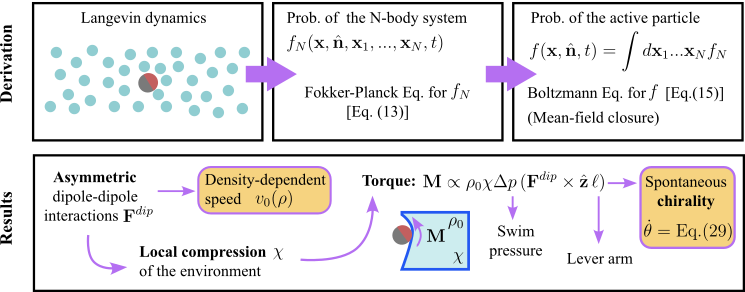}
	\caption{\textbf{Theory for the motion of an active particle in compressible environments.} Schematic representation of the key steps and results described in the Theoretical Section. The combination of asymmetric interactions (between the Janus active particle and the passive spheres) and a compressible matrix generates a torque leading to a spontaneous chiralization of the active motion. 
	}
	\label{fig6}
\end{figure*}

The dimensionless parameter i) is determined from the radius of the experimental colloids. The dimensionless parameters ii) and iii) can be calculated using the Einstein relations for the translational and rotational diffusion coefficients in water solutions. In dimensional units, the translational diffusion coefficient of passive and active particles, $\rm D_p$ and $\rm D_a$, can be obtained by using the following formula $\rm \xi D = k_B T$, where $\rm k_B$ is Boltzmann's constant and $\rm \xi =6\pi \eta_0 R$ denotes the translational friction coefficient. Here, $\eta_0$ represents the water viscosity, $\rm T$ is the ambient temperature, while $\rm R$ denotes the particle radius. By choosing $\rm R_p$ or $\rm R_a$ in the expression for $\xi$, we obtain $\rm \xi_p$ or $\rm \xi_a$, respectively. In this way, passive and active particles evolve with the following reduced diffusion coefficients, $\rm D_p\xi_p/\epsilon=0.01$ and $\rm D_a=D_p R_a^2/R_p^2$. The rotational friction coefficient is given by $\rm \xi_R=8 \pi \eta_0 R_a^3$ while the rotational diffusion coefficient is fixed by the relation $\rm R_a^2 D_R=D_a \,3/4$. The two remaining parameters, iv) and v) can be fitted by comparing simulations with experiments for every combination of packing fractions and electric fields. The parameter $\rm K/\epsilon$ is determined through simulations of purely passive particles following the dynamics in Eq.~\eqref{eq:passive_dynamics}. The value of $\rm K/\epsilon$ is iteratively changed until the pair correlation function $\rm g(r)$ obtained in simulations aligns with the experimental curve. This value is used as the strength of active-passive dipolar interactions. Similarly, $f_{\rm s} {\rm R_p} /\epsilon$ is obtained by evolving the dynamics in Eq.~\eqref{eq:activedynamics} for an active particle in a passive environment, iteratively repeating the numerical study until the active particle velocity $\dot{\mathbf{x}}$ aligns with the experimental result.

\section{Theoretical Section}\label{TheoSec}

In the following we provide a detailed theory supporting the experimental observation of a self-sustained reorientation of active particles when they cruise in compressible colloidal monolayers. A visual guide through the theory is given in Fig.~\ref{fig6}.

\subsection{Fokker-Planck equation for an active particle immersed in a passive bath}

An active particle in a passive bath is described by coupled stochastic differential equations: Eq.~\eqref{eq:passive_dynamics} models passive particles, while Eqs.~\eqref{eq:activedynamics} describe the evolution of position and orientation for the self-propelled colloid. The dynamics for the orientational angle $\theta$ can be expressed in polar coordinates keeping the Ito convention for interpreting the stochastic noise as follows:
\begin{equation}
	\dot{\boldsymbol{\rm n}} =-{\rm D_R}\boldsymbol{\rm n}+ \sqrt{2 \rm D_R}  \boldsymbol{\chi}^{\rm a} +\frac{1}{\xi_{\rm R}} \mathbf{M}\times \boldsymbol{\rm n} \,.
	\label{Cdynamicequation3}
\end{equation}
Here, the noise vector $\rm \boldsymbol{\chi}^a=(0, 0, \xi^r_z)$ is a noise vector and torque $\mathbf{M}$ are directed along $\hat{\mathbf{z}}$, {\sl i.e.} normally to the plane of motion. The latter term depends on the position of the environmental particles as $\mathbf{M}=\boldsymbol{\ell} \times \mathbf{F}^{\rm eff} = \hat{\mathbf{n}}\ell \times \sum_j \mathbf{F}^{\rm dip}_{j}$. Switching to the Fokker-Planck equation for the  probability distribution $f_{\rm N}(\mathbf{x},\hat{\mathbf{n}}, \mathbf{x}_1, ...,  \mathbf{x}_{\rm N}, \rm t)$ of the full many-body interacting system, we obtain
\begin{flalign}
	\label{eq:fullfokkerplanck}
	\frac{\partial}{\partial \rm t} &f_{\rm N} =\nabla_{i} \left( \rm  D_p \nabla_i - \frac{\mathbf{F}_{i}^{a-p}}{\xi_p}  \right) f_{\rm N} + \nabla \left(\rm D_a \nabla - \frac{\mathbf{F}_{\rm a}^{\rm a-p}}{\xi_a} \right) f_{\rm N} \nonumber\\
	&- v_0\hat{\mathbf{n}} \cdot\nabla f_{\rm N}  
	+ \rm D_R \mathcal{L}_a f_N - \frac{1}{\xi_R} \frac{\partial}{\partial \hat{\mathbf{n}}} \cdot\left[  \mathbf{M}\times \hat{\mathbf{n}}\right] f_N \,.
\end{flalign}
Here, $\nabla$ and $\nabla_i$ denote the derivative with respect to active particle position $\mathbf{x}$ and the passive particle position $\mathbf{x}_i$, respectively. In addition, we have defined the active speed $v_0=f_{\rm s}/\xi_{\rm a}$ and we remind that the total force acting on the active particle due to the passive particles can be decomposed as $\mathbf{F}^{\rm a-p}_a=\mathbf{F}^{\rm WCA}_{\rm a}+ \mathbf{F}_{\rm a}^{\rm dip}$ being due to steric repulsions (WCA potential) and dipolar potentials. The latter additionally exerts a torque on the dynamics of the orientational angle. By contrast, the total force acting on each passive particle is due to the active colloid and other passive particles, $\mathbf{F}_i^{\rm a-p}$ and $\mathbf{F}_i^{\rm p-p}$, respectively. The operator $\mathcal{L}_a$ accounts for the dynamics of the active force and has the following form~\cite{marconi2021hydrodynamics}:
\begin{equation} \label{eq:ActiveForce_evolutiongeneral}
	\mathcal{L}_{\rm a} f
	=\frac{\partial }{\partial \hat{\mathbf{n}}} \cdot\Bigl(  \hat{\mathbf{n}} +  \frac{\partial }{\partial \hat{\mathbf{n}}}\cdot 
	\boldsymbol{\mathcal{D}}\Bigl)f \,,
\end{equation}
where the matrix $\boldsymbol{\mathcal{D}}$ has the following spatial components:
\begin{equation}
	\boldsymbol{\mathcal{D}} =
	\left( \begin{array}{cccc}
		n_y^2  & -n_x n_y  \\ 
		n_yn_x & n_x^2 
	\end{array} \right)\,.\nonumber\\
	\label{D_ABP}
\end{equation}
$\mathcal{L}_{\rm a}$ is formed by two terms: a ``deterministic'' drag term proportional to the first derivative with respect to $\hat{\mathbf{n}}$  and a ``diffusive'' term proportional to the second derivative with respect to $\hat{\mathbf{n}}$. Moreover, we remark the following property, which will be used later: 
\begin{equation*}
	\langle \boldsymbol{\mathcal{D}}\rangle  =   {\bf I} \,,
\end{equation*}
where $\mathbf{I}$ is the identify matrix and the average is performed over all the variables of the system. The approximation above does not affect the coarse-grained description~\cite{marconi2021hydrodynamics}.

\subsection{From the Fokker-Planck to the Boltzmann equation}

By integrating the Fokker-Planck equation over the environment degree of freedom, we obtain an effective Boltzmann equation for the system under investigation. Specifically, by defining the single-body probability distribution $f= f(\mathbf{x}, \hat{\mathbf{n}}) = \int d\mathbf{x}_1 ... d\mathbf{x}_{\rm N} f_{\rm N}(\mathbf{x}, \hat{\mathbf{n}}, \mathbf{x}_1 ... \mathbf{x}_{\rm N})$ and integrating the Fokker-Planck equation over $\mathbf{x}_1, ...,  \mathbf{x}_{\rm N}$, we obtain
\begin{flalign}
	&\frac{\partial}{\partial \rm t} f = {\rm D_a}\nabla^2 f 
	- \nabla\cdot\int d\mathbf{x}' \frac{\mathbf{F}_{\rm a}^{\rm a-p}(|\mathbf{x}-\mathbf{x}'|)}{\xi_{\rm a}}  f_2 
	- v_0\hat{\mathbf{n}}\cdot \nabla f  \nonumber\\ 
	&+ {\rm D_R} \mathcal{L}_{\rm a} f  
	- \frac{\partial}{\partial \hat{\mathbf{n}}}\cdot  \int d\mathbf{x}' f_2 \mathbf{M}(|\mathbf{x}-\mathbf{x}'-\boldsymbol{\ell}|)\times  \frac{\hat{\mathbf{n}}}{\xi_{\rm R}} \,.
\end{flalign}
This equation for $f$ involves the two body probability distribution $f_2=f_2(\mathbf{x}, \mathbf{x}', \hat{\mathbf{n}}, \rm t)$ depending on the coordinates of one active particle and one passive particle. To proceed further, it is necessary to express $f_2$ in terms of $f$ finding a suitable closure for the BBGKY hierarchy. As shown below, the first term involving $\mathbf{F}^{\rm a-p}$ can be treated as in Ref.~\cite{speck2015dynamical} and leads to an effective active speed $v_0 \to v_0(\rho_{\rm env})$ which depends on the passive particle density, {\sl i.e.} on how many interactions take place on the active particle. The second term involving $\mathbf{M}$ needs a different treatment which will be responsible for the chiral motion observed in the active particle dynamics.

\subsection{Approximation to close the BBGKY hierarchy}

We approximate the two-body probability distribution $f_2(\mathbf{x}, \mathbf{x}', \hat{\mathbf{n}}, t)$ depending on the active and passive particle coordinates, $\mathbf{x}$ and $\mathbf{x}'$, as
\begin{equation}
	f_2(\mathbf{x}, \mathbf{x}', \mathbf{n}, {\rm t}) = \rho_{\rm env}(\mathbf{x}') f (\mathbf{x}, \mathbf{n}, t) {\rm g}
\end{equation}
where
\begin{equation}
	g=\begin{cases}
		{\rm g}(|\mathbf{x}-\mathbf{x}'|, \varphi, {\rm t})\\
		{\rm g_m}(|\mathbf{x}-\mathbf{x}' - \hat{\mathbf{n}}\ell|, \varphi, {\rm t})\,.
	\end{cases}
\end{equation}
Here, we have neglected the time evolution of $\rho_{\rm env}(\mathbf{x}', {\rm t}) = \rho_{\rm env}(\mathbf{x}')$ which is supposed to relax fast compared to the active particle density. The term ${\rm g}(|\mathbf{x}-\mathbf{x}'|, \varphi, {\rm t})$ is the pair correlation function which depends on the relative distance between active and passive particles as well as the angle $\varphi=\varphi(\theta, {\rm t})$ enclosed by the displacement vector $\mathbf{x}-\mathbf{x}'$ and the active particle orientation $\hat{\mathbf{n}}=(\cos\theta, \sin\theta)$. $\rm g_m$ is the same pair correlation function evaluated at the distance between the center of the passive particle and the center of the active particle cap -- which does not coincide with the geometric center. This approach resembles the one developed in Ref.~\cite{speck2015dynamical}, with the difference that the active particle interacts only with the surrounding passive particles. Under this assumption, we estimate the interaction term as
\begin{flalign}
	&\frac{1}{\xi_{\rm a}}\nabla \cdot\mathbf{F}=\nabla \cdot\int d\mathbf{x}' \frac{\mathbf{F}_{\rm a}^{\rm a-p}(|\mathbf{x}-\mathbf{x}'|)}{\xi_{\rm a}}  f_2  \nonumber\\
	 &=- \frac{1}{\xi_{\rm a}}\nabla  f \int d\mathbf{x}' u'(|\mathbf{x}-\mathbf{x}'|) \frac{\mathbf{x}-\mathbf{x}'}{|\mathbf{x}-\mathbf{x}'|}  \rho_{\rm env}(\mathbf{x}') g \,,
\end{flalign}
where we have introduced the mean force $\mathbf{F}$ as the integral of the interactions. The interaction force $\mathbf{F}^{\rm a-p}$ is the sum of the force due to WCA potentials and dipolar interactions. Here, we have neglected the asymmetry in the dipolar interactions. Specifically, we have assumed $\ell \approx 0$ since the asymmetry does not fundamentally contribute to the translational dynamics of the active particle. 

To proceed, we decompose the mean force as 
\begin{equation}
	\mathbf{F} = (\hat{\mathbf{n}}\cdot \mathbf{F}) \hat{\mathbf{n}} + \delta \mathbf{F} =  (\hat{\mathbf{n}}\cdot \mathbf{F}) \hat{\mathbf{n}} + \frac{\rm D_a}{\rm D_R} F_{\parallel} \nabla f
\end{equation}
where 
\begin{equation}
	F_{\parallel} = \frac{\rm D_R}{\rm D_a} \left[ \frac{\nabla f - (\hat{\mathbf{n}} \cdot \nabla f) \hat{\mathbf{n}}}{|\nabla f|^2}  \right] \cdot \mathbf{F}\,.
\end{equation}
Since $\delta \mathbf{F}$ is small, we can immediately find the leading contribution by scalarly multiplying the previous expression by $\hat{\mathbf{n}}$, obtaining
\begin{flalign}
		\frac{1}{\xi_{\rm a}}\mathbf{F} \cdot \hat{\mathbf{n}} &\approx 
		- \frac{1}{\xi_{\rm a}}  f \int d\mathbf{x}' U'(|\mathbf{x}-\mathbf{x}'|)  
		\rho_{\rm env}(\mathbf{x}') {\rm g} \cos{\varphi} \nonumber\\
		&= - \frac{1}{\xi_{\rm a}}  f  \tilde{\rho}_{\rm env} \int dr' r'  U'(r')  \int_0^{2\pi} d\varphi \,{\rm g}(r', \varphi, {\rm t}) \cos{\varphi} \nonumber\\ 
		&= - \frac{1}{\xi_{\rm a}}  f (\mathbf{x},\hat{\mathbf{n}}, {\rm t})  \tilde{\rho}_{\rm env} \zeta \,.
\end{flalign}
In the first equality, we have used that $\mathbf{F}^{\rm a-p} = -\nabla \rm U$ and $\rm U$ is the total interaction potential comprising the dipolar interactions and the WCA potential. In the last equality, we have considered a uniform passive environment $\rho(\mathbf{x}')\approx \tilde{\rho}$, switched to polar coordinates and introduced the coefficient $\zeta$ as the result of the radial and angular integral reported above. Note that if ${\rm g}(r', \varphi)$ does not depend on $\varphi$ as in the absence of activity, then $\zeta=0$. This implies that the interaction term reads
\begin{equation}
	\frac{1}{\xi_{\rm a}}\nabla \cdot\mathbf{F}= - \frac{1}{\xi_{\rm a}}   \tilde{\rho}_{\rm env} \zeta\nabla f  - \frac{F_{\parallel}}{\xi_{\rm a}} \frac{\rm D_a}{\rm D_R} \nabla^2 f \,.
\end{equation}
Assuming that both $\zeta$ and $F_{\parallel}$ are constant terms, the Boltzmann equation takes the following form:
\begin{flalign} \label{eq:Boltzmann1} 
	\frac{\partial}{\partial {\rm t}} f = \nabla \cdot \left( {\rm D_{eff}}\nabla  -  v(\rho)\hat{\mathbf{n}}\right) f  + {\rm D_R} \mathcal{L}_{\rm a} f  \\ 
	-\frac{1}{\xi_{\rm R}}\frac{\partial}{\partial \hat{\mathbf{n}}}\cdot \left[ \int d\mathbf{x}'\mathbf{M}(|\mathbf{x}-\mathbf{x}'|)f_2\right]\times \hat{\mathbf{n}} \nonumber\,
\end{flalign}
where 
\begin{subequations}
	\label{eq:v0andDeff}
	\begin{flalign}
		&v(\rho) = v_0 - \tilde{\rho}_{\rm env} \frac{\zeta}{\xi_{\rm a}}\\
		&D_{\rm eff}={\rm D_a} - \frac{F_{\parallel}}{\xi_{\rm a}} \frac{\rm D_a}{\rm D_R} \,.
	\end{flalign}
\end{subequations}
As a consequence, the environmental particles have two effects on the effective dynamics of the active particle: (i) the effective swim velocity is reduced as the density of the environment is increased and (ii) the effective diffusion coefficient of the active particle is decreased.

The approximation of the torque term is more problematic. A net torque on the active particle naturally arises only because the dipolar force is no longer applied in the geometric center of the particle $\mathbf{x}$ but rather in the position $\mathbf{x}-\boldsymbol{\ell}$, where $\boldsymbol{\ell}=\hat{\mathbf{n}}\ell$. By introducing the average torque, $\boldsymbol{\mathcal{M}}$, this term can be expressed as
\begin{flalign}
		&\frac{\partial}{\partial \hat{\mathbf{n}}} \boldsymbol{\mathcal{M}}\times \frac{\hat{\mathbf{n}}}{\xi_{\rm R}} 
		=\frac{\partial}{\partial \hat{\mathbf{n}}}  \int d\mathbf{x}' f_2\mathbf{M}(|\mathbf{x}-\mathbf{x}'-\boldsymbol{\ell}|) \times \frac{\hat{\mathbf{n}}}{\xi_{\rm R}} \\
		&= - \frac{\partial}{\partial \hat{\mathbf{n}}}  f  \int d\mathbf{x}' \rho_{\rm env}(\mathbf{x}') {\rm g_m}\left\{\mathbf{F}^{\rm dip}(|\mathbf{x}- \mathbf{x}'-\boldsymbol{\ell}|) \times \boldsymbol{\ell}\right\} \times  \frac{\hat{\mathbf{n}}}{\xi_{\rm R}}  \nonumber
		  \,.
\end{flalign}
Manipulating the latter term is hard. However, we can estimate the average torque by considering that it is generated by a density gradient in the environment. This density gradient emerges because the dipolar interactions are applied in the center of the uncoated hemisphere. Therefore, dipolar interactions induce a density environmental change compared to the the homogeneous value $\rho(\mathbf{x}') = \rho_0 + \lambda\Delta \rho (\mathbf{x}')$. Here, the length $\lambda$ represents the typical distance where the density change takes place. The average torque $\boldsymbol{\mathcal{M}}$ becomes 
\begin{flalign}
	\boldsymbol{\mathcal{M}}&=\int d\mathbf{x}' \left\{\mathbf{F}^{\rm dip}(|\mathbf{x}-\boldsymbol{\ell}- \mathbf{x}'|) \times \boldsymbol{\ell}\right\} \rho_{\rm env}(\mathbf{x}') {\rm g_m}\nonumber\\
&	\approx \Delta \rho \pi \lambda^2 \,\tilde{\rm g}\, |\tilde{\mathbf{F}}^{\rm dip} | \ell \,\hat{\mathbf{z}}
\end{flalign}
where $\tilde{\mathbf{F}}^{\rm dip} = \mathbf{F}^{\rm dip}(\rm R_p+R_a)$ is the dipolar force calculated at the distance between passive and active particle and $\tilde{\rm g}=\int d\varphi {\rm g_m}(\rm R_p + R_a, \varphi)$ is the pair correlation function evaluated at the contact point and averaged over the polar angle $\varphi$. Note that this approximation is equivalent of considering the $\delta$-Dirac function approximation for the force. We can proceed further and estimate $\Delta \rho$ by treating the active force due to active particle as a simple external force acting on the passive environment:
\begin{equation}
	\label{eq:app_eq_Deltarho}
	\Delta \rho = -\rho_0 \frac{\Delta \rm V}{\rm V} = \rho_0 \chi \Delta {\rm p} =\rho_0 \chi \frac{\gamma v_0}{2\pi \lambda}
\end{equation}
being $\chi$ the compressibility of the passive environment and $\Delta \rm p$ the pressure difference. $\Delta \rm p$ is estimated in the last equality of Eq.~\eqref{eq:app_eq_Deltarho} as the swim pressure generated by the active velocity.
Consequently, the torque term is approximated as
\begin{equation}
	\frac{1}{\xi_{\rm R}}\frac{\partial}{\partial \hat{\mathbf{n}}}\cdot \boldsymbol{\mathcal{M}}\times \hat{\mathbf{n}} \approx \langle \dot{\theta} \rangle \, \frac{\partial}{\partial \hat{\mathbf{n}}}\cdot\left(\hat{\mathbf{z}}\times \hat{\mathbf{n}}\right) \, f (\mathbf{x}, \hat{\mathbf{n}}, t)
\end{equation}
where we have introduced the typical angular velocity $\langle \dot{\theta} \rangle$ as
\begin{equation}
	\label{eq:prediction}
	\langle \dot{\theta} \rangle= \rho_0 \chi \frac{\gamma v_0}{2 \xi_{\rm R}}  \lambda \,\tilde{\rm g}\, |\tilde{\mathbf{F}}^{\rm dip} | \ell \,.
\end{equation}
This angular velocity vanishes in the absence of activity $v_0=0$, in the absence of dipolar interactions or if these interactions are entirely applied in the geometric center ($\ell =0$). Eq.~\eqref{eq:prediction} predicts the scaling of $\langle \dot{\theta} \rangle$ as a function of $v_0$ reported in experiments and numerical simulations.

By replacing the expression for $\boldsymbol{\mathcal{M}}$ in Eq.~\eqref{eq:Boltzmann1}, we obtain
\begin{equation}
	\label{eq:Boltzmann2}
	{\partial_t} f = \nabla \cdot \left( {\rm D_{eff}}\nabla  -  v(\rho)\hat{\mathbf{n}}\right) f  + {\rm D_R} \mathcal{L}_{\rm a} f  - \langle \dot{\theta} \rangle\frac{\partial}{\partial \hat{\mathbf{n}}}\cdot \left[\hat{\mathbf{z}}\times \frac{\hat{\mathbf{n}}}{\xi_{\rm R}}\right]f \,.
\end{equation}
This is the effective Boltzmann-like equation for a non-interacting chiral active particle. Here, the interactions with the passive particle of the environment induced a density-dependent cruising velocity $v(\rho)$ and an effective torque which gives rise to a chirality $\langle \dot{\theta} \rangle$.

\subsection{Derivation of an effective Fick's equation}

To derive a Fick's equation, we introduce the density and polarization fields, which depend on the active particle position $\mathbf{x}$ and time $\rm t$. The local density, $\rho=\rho(\mathbf{x},t)$, is defined by integrating the single-body probability distribution $f$ over the self-propulsion vector:
\begin{equation}
		\rho(\mathbf{x}, {\rm t})= \int d\mathbf{n} \, f(\mathbf{x}, \hat{\mathbf{n}}, {\rm t}) \,.
		\label{density}
\end{equation}
On the other hand, the local polarization field, $\mathbf{p}=\mathbf{p}(\mathbf{x},{\rm t})$, can be introduced by multiplying the single-body distribution $f$ by $\mathbf{n}$ and then integrating over $\mathbf{n}$: 
\begin{equation}
		\mathbf{p}(\mathbf{x}, {\rm t})=\int d\mathbf{n} \, f(\mathbf{x}, \hat{\mathbf{n}}, {\rm t})\,\hat{\mathbf{n}}  \,.
		\label{definizionepolarizzazione}
\end{equation}
By integrating the closed Boltzmann equation for the single-body probability distribution $f$ over the self-propulsion vector $\hat{\mathbf{n}}$ (Eq.~\eqref{eq:Boltzmann2}), we obtain the time evolution for the density field $\rho(\mathbf{x}, {\rm t})$:
\begin{equation}
	\label{eq:densityeq}
	\frac{\partial}{\partial {\rm t}} \rho = - \nabla \cdot \left[ v(\rho) \mathbf{p} -  {\rm D_{eff}} \nabla \rho \right] \,,
\end{equation}
where $v(\rho)$ and $\rm D_{eff}$ are given by Eqs.~\eqref{eq:v0andDeff}.
By multiplying Eq.~\eqref{eq:Boltzmann2} by $\hat{\mathbf{n}}$ and then integrating over $\hat{\mathbf{n}}$, we derive the polarization balance equation, which reads
\begin{equation}
	\label{eq:polarization}
	\frac{\partial}{\partial {\rm t}} \mathbf{p} = -\frac{1}{2}\nabla\left( v(\rho) \rho \right) + {\rm D_{eff}} \nabla^2 \mathbf{p} - {\rm D_R} \mathbf{p} +\langle \dot{\theta} \rangle\hat{\mathbf{z}}\times \mathbf{p} \,.
\end{equation}
Here, we have closed the coarse-grained equations by approximating the quadrupolar tensor $\langle \hat{\mathbf{n}}\hat{\mathbf{n}}\rangle=\mathbf{I}/2$. The polarization relaxes faster than the density. Thus we can neglect time and spatial derivatives in Eq.~\eqref{eq:polarization}, obtaining
\begin{flalign} 
	\mathbf{p}&= -\frac{1}{2\rm D_R} \nabla (v(\rho) \rho)  \left( \mathbf{I} - \boldsymbol{\epsilon} \langle \dot{\theta} \rangle \right)^{-1}  \\
	&= -\frac{1}{2\rm D_R (1+\langle \dot{\theta} \rangle^2)} \nabla (v(\rho) \rho)  \left( \mathbf{I} + \boldsymbol{\epsilon} \langle \dot{\theta} \rangle \right)
\label{eq:steadystatepolarization}
\end{flalign}
where the antisymmetric (2D Levi-Civita) tensor $\boldsymbol{\epsilon}$ has element $\epsilon_{xx}=\epsilon_{yy}=0$ and $\epsilon_{xy}=-\epsilon_{yy} = 1$. The solution for the steady-state polarization field $\mathbf{p}(\mathbf{x})$ admits transverse gradients that explains the helical trajectory experimentally and theoretically obtained.
Indeed, by replacing the solution of Eq.\eqref{eq:steadystatepolarization} in the equation for $\rho$ (Eq.\eqref{eq:densityeq}), we obtain
\begin{equation}
	\frac{\partial}{\partial t} \rho =  {\rm D_{eff}} \nabla^2 \rho +  \frac{ \left( \mathbf{I} + \boldsymbol{\epsilon} \langle \dot{\theta} \rangle \right)}{2 {\rm D_R} (1+\langle \dot{\theta} \rangle^2)} \nabla \cdot \left[ v(\rho)\nabla (v(\rho) \rho)\right] \,.
\end{equation}
The antisymmetric structure of the tensor $\boldsymbol{\epsilon}$ implies that the system shows odd diffusivity. This is a well-known consequence of the presence of chirality as observed in previous studies~\cite{hargus2021odd}.

\section*{Supporting Information}
Supplementary information is available for this paper.

\section*{Acknowledgments}
	Financial support by the Deutsche Forschungsgemeinschaft (DFG) is gratefully acknowledged. TP and VMSGT acknowledge funding within project PA 459/18-2. IB acknowledges funding within project BU 4040/3-1. HL acknowledges funding within project LO 418/29-1. IM acknowledges support from the Alexander von Humboldt Foundation. LC acknowledges funding from the Italian Ministero dell'Universit\'a e della Ricerca under the programme PRIN 2022 (``re-ranking of the final lists''), number 2022KWTEB7, cup B53C24006470006.

\section*{Conflict of Interest}
The authors declare no competing interests.

\section*{Data Availability Statement}
Experimental and numerical trajectories of active particles are freely available as a supplement to this manuscript at https://github.com/isham612/Experimental-and-Numerical-Trajectories/tree/main. The other data are available from the corresponding authors upon reasonable request.

\section*{Authors contribution}
Conceptualization: HL, TP, IB. Experiments: VMSGT. Numerical simulations: IM. Theory: LC, HL. Data analysis: VMSGT, IM, IB. Supervision: HL, TP, IB. Writing original draft: LC, IB. Writing review and editing: VMSGT, IM, LC, HL, TP, IB.

\bibliographystyle{naturemag}

\end{document}